\documentclass[10pt, pra, aps, nopacs, twocolumn, longbibliography, floatfix,superscriptaddress,amsmath]{revtex4-2}
\usepackage{graphicx}
\usepackage{wrapfig}

\usepackage{braket}
\usepackage{mathtools}
\usepackage[titletoc]{appendix}
\usepackage{appendix}
\usepackage{float}
\usepackage{bm}
\usepackage{amssymb}
\usepackage{siunitx}
\DeclareSIUnit\TunnelingEnergy{J}

\usepackage[dvipsnames]{xcolor}
\definecolor{darkblue}{rgb}{0.0, 0.0, 0.75}

\usepackage{color}
\usepackage[colorlinks=true,
            linkcolor=darkblue,
            urlcolor=darkblue,
            citecolor=darkblue]{hyperref}

	\usepackage[normalem]{ulem} 
	\definecolor{mgreen}{RGB}{1,123,0}

\def \br{{\bf r}}

\def \mb{\mathrm{b}}

\def \m2D{\mathrm{2D}}

\def \mt{\mathrm{t}}
\def \mms{\mathrm{ms}}

\def \mG{\mathrm{G}}

\def \mb{\mathrm{b}}

\def \mms{\mathrm{mm/s}}

\def \br{\mathbf{r} }

\DeclareMathOperator{\sig}{sig}

\begin{document}
\title{Realizing an Atomtronic AQUID in a Rotating-Box Potential}
\author{Kaspar Görg}
\affiliation{Zentrum f\"ur Optische Quantentechnologien and Institut f\"ur Quantenphysik, Universit\"at Hamburg, 22761 Hamburg, Germany}
\author{Ludwig Mathey}
\affiliation{Zentrum f\"ur Optische Quantentechnologien and Institut f\"ur Quantenphysik, Universit\"at Hamburg, 22761 Hamburg, Germany}
\affiliation{The Hamburg Centre for Ultrafast Imaging, Luruper Chaussee 149, Hamburg 22761, Germany}
\author{Vijay Pal Singh}
\affiliation{Quantum Research Centre, Technology Innovation Institute, Abu Dhabi, UAE}
\date{\today}
%
%
\begin{abstract}
%
%
Atomtronic devices are matter-wave circuits designed to emulate the functional behavior of their electronic counterparts.
Motivated by superconducting quantum interference devices (SQUIDs), 
atomic quantum interference devices (AQUIDs) have been developed using Bose-Einstein condensates (BECs) confined in toroidal geometries. 
Here, we propose and numerically investigate an alternative implementation of an AQUID based on a BEC confined in a rotating box potential. A ring-like topology is established by introducing a central depletion region via a repulsive potential barrier. We observe the hallmark AQUID feature---quantized phase winding that increases in discrete steps with angular velocity. Centrifugal effects induced by rotation degrade phase coherence and impair AQUID performance, 
which we mitigate by applying a counteracting harmonic confinement. 
Phase slips are found to be mediated by a vortex propagating from the central depletion zone to the edge of the condensate. To characterize the voltage response, we induce a bias current by translating the box along its long axis while keeping the central barrier fixed. This generates a density imbalance between the two reservoirs, 
exhibiting a periodic dependence on angular velocity---analogous to the voltage-flux relation in electronic SQUIDs. 
Our results demonstrate that rotating box geometries provide a viable and flexible platform for realizing atomtronic AQUIDs with controllable dynamics and well-defined response characteristics. 
\end{abstract}

\maketitle
%

\section{Introduction}
Atomtronics utilizes the analogy between Cooper pairs in superconductors 
and bosonic atoms in a Bose-Einstein condensate (BEC) to enable the design of matter-wave circuits that mimic behavior of their electronic counterparts \cite{Amico2021, Amico2022}.
A wide variety of atom-based components, including diodes, transistors, and batteries, have been theoretically proposed and experimentally realized \cite{AtomtronicsFirstPaper, Amico2014, Micheli2004, Stickney2007, Ruschhaupt2007, Thorn2008, Pepino2009, Krinner2017, Zozulya2013, Caliga2017, Caliga2016, Jahrling2024}. 
Among these, atomic Josephson junctions (JJs) play a central role, giving rise to 
hallmark phenomena such as 
dc-ac Josephson effects \cite{Albiez2005, Levy2007}, current-phase relations \cite{Luick2020, Kwon2020, Pace2021}, and even Shapiro steps \cite{SinghShapiro, Del_Pace2024,Bernhart2024} and parametric amplification \cite{Singh2025} in driven setups.

A particularly compelling example is the superconducting quantum interference device (SQUID), 
which comprises two JJs in a superconducting loop. 
In such systems, the coupling of Cooper pairs to the electromagnetic vector potential leads to flux quantization, which underpins the SQUID's exceptional sensitivity to magnetic fields \cite{gross2016, vrba2006}. 
In the atomtronic analog, a BEC confined to a ring-shaped trap plays the role of the loop, 
and narrow potential barriers serve as the Josephson junctions. 
Rotating the barriers or the entire trap introduces a synthetic gauge field, 
emulating magnetic flux and making the system sensitive to angular velocity instead of magnetic fields. 
This analogy is the basis for the development of atomic quantum interference devices (AQUIDs) with potential application as rotation sensors. 
Ring-shaped condensates exhibit persistent superfluid currents due to quantized angular momentum \cite{Ramanathan2011, Pace2022, Cai2022, Mathey2014, POLO2025}. 
Multiple toroidal atomtronic AQUIDs have been demonstrated, both experimentally \cite{Wright2013, ryu2013, Eckel2014, Campbell2014, ryu2020} and theoretically \cite{mathey2016, kiehn2022}, 
revealing key features such as quantized phase winding, phase slips, and flux-dependent interference patterns.

In this work, we extend the AQUID concept beyond toroidal geometries by creating an atomtronic AQUID in a rotating box potential. Our setup consists of a rectangular box trap containing a BEC and a central repulsive barrier  
that effectively divides the condensate into two reservoirs connected by two weak links, acting as Josephson junctions. 
By rotating the box, we introduce a synthetic gauge field that emulates a magnetic field, giving rise to screening currents, quantized circulation, and phase-slip events. The resulting stepwise evolution of the global phase winding with angular velocity confirms the AQUID response. 
To probe the system's imbalance-flux characteristics, we laterally translate the box potential relative to the fixed central barrier, thereby introducing a density imbalance that mimics a bias current across an electronic SQUID. 
We observe clear modulation of the density imbalance with angular velocity, 
consistent with interference effects between the two junctions. 
For optimal barrier parameters, the system exhibits near-ideal AQUID behavior, 
characterized by regular phase-winding step spacing and pronounced imbalance modulation near half-integer flux quanta.  
All results are obtained using classical-field simulations, which capture condensate dynamics beyond the mean-field approximation \cite{Blakie2008, Polkovnikov2010, Singh2017}.


This paper is structured as follows. 
In Sec. \ref{sec:simulation_method}, we describe the classical-field simulation method used to model condensate dynamics. 
In Sec. \ref{section:SQUIDImplementation}, we present the protocol for creating an atomtronic AQUID in a rotating box potential. 
In Sec. \ref{section:phaseSlip}, we analyze and optimize the phase-slip dynamics in the zero-voltage state. 
In Sec. \ref{sec:time_phase_slip}, we investigate the vortex dynamics associated with a single phase-slip event. 
In Sec. \ref{sec:voltage_state}, we examine the voltage-state behavior by studying the dependence of the density imbalance on angular velocity. 
Finally, we conclude in Sec. \ref{sec:con}.


%
\section{Simulation method}\label{sec:simulation_method}
We employ a semiclassical approach based on classical-field dynamics within the truncated Wigner approximation, which is a well-established framework for modeling condensate dynamics in ultracold atom experiments \cite{Blakie2008, Polkovnikov2010, Singh2017, Singh2021}. 
Importantly, this method includes thermal and quantum fluctuations, 
enabling us to assess the robustness of our protocols under realistic, finite temperature conditions. 
The system is described by the Hamiltonian 
\begin{multline}\label{eq:hamiltonian}
    \hat{H}_0 = \int \mathrm{d}\mathbf{r} \Bigl[ 
     \frac{\hbar^2}{2m} \nabla \hat{\psi}^\dagger(\mathbf{r}) \cdot \nabla \hat{\psi}(\mathbf{r}) 
    + V(\mathbf{r}) \hat{\psi}^\dagger(\mathbf{r}) \hat{\psi}(\mathbf{r})\\
    + \frac{g}{2} \hat{\psi}^\dagger(\mathbf{r}) \hat{\psi}^\dagger(\mathbf{r}) \hat{\psi}(\mathbf{r}) \hat{\psi}(\mathbf{r})
    \Bigr],
\end{multline}
where $\hat{\psi}$ and $\hat{\psi}^\dagger$ are the bosonic annihilation and creation operators. 
The 2D interaction parameter $g = \tilde{g} \hbar^2/m$ is given in terms of the dimensionless interaction strength $\tilde{g}=\sqrt{8\pi}a_s/l_z$, where $\hbar$ is the reduced Planck constant, $m$ is the atomic mass, $a_s$ is the $s$-wave scattering length, and $l_z$ is the harmonic oscillator length in the transverse direction.
The external potential $V(\mathbf{r})$ consists of a box potential and a repulsive Gaussian potential, 
which are described in detail in the next section. 
We consider a homogeneous cloud confined in a rectangular box of dimensions $L_x \times L_y = \SI{128}{\micro\meter} \times \SI{42}{\micro\meter}$, with density  $n_0=\SI{1.2}{\per\micro\meter\squared}$ and $\tilde{g}=0.5$. These are typical parameters for experiments with $^6\mathrm{Li}_2$ molecules \cite{Luick2020}.
For numerical simulations, we discretize real space on a $135 \times 135$ lattice with spacing $l = \SI{1}{\micro\meter}$, which is sufficiently large for the box to rotate within the lattice boundaries.

In the classical-field approximation, the field operators $\hat{\psi}$ are replaced by complex numbers $\psi$ in both the Hamiltonian and the equations of motion. 
We sample the initial states from a grand canonical ensemble at fixed temperature $T$ and chemical potential $\mu$ using a classical Metropolis algorithm. 
The resulting distribution includes fluctuations of $\psi(\br, t=0)$ around its mean-field value. 
Each initial state is then evolved in time using the equations of motion 
\begin{align}\label{eq:dscrt_GPE}
i \hbar \dot{\psi}_i = - J \sum_{j} \psi_{j(i)} + (V_i + U |\psi_{i}|^2) \psi_i, 
\end{align}
where $i$ denotes the lattice site and $j$ runs over nearest neighbors. 
$J = \hbar^2/(2ml^2)$ is the tunneling energy, and $U=g/l^2$ is the onsite interaction strength. 
For our system of $^6\mathrm{Li}_2$ molecules, the tunneling energy is $J/k_\mathrm{B}= 20.2\, \mathrm{nK}$. 
We perform the time evolution using the fourth-order Runge-Kutta-Cash-Karp method 
with adaptive time step size \cite{cash1990}.
To reduce computational overhead, the wavefunction at sites more than \SI{5}{\micro\meter} away from the box boundary is kept constant and excluded from the time evolution. 
A condensate fraction of about $90\%$ is ensured by choosing a low temperature $T$ in the initial state sampling.

%
\section{AQUID Protocol}\label{section:SQUIDImplementation}
In this section, we describe our protocol to implement the atomtronic AQUID. 
We consider a box potential of the form
\begin{multline}
	V_\mb = V_{\mb,0} \Bigl[ 1- \sig{\Big( \frac{x_\mb}{s} \Big)}
    \sig{\Big(\frac{L_x-x_\mb}{s}\Big)} \\
    \sig{\Big(\frac{y_\mb}{s}\Big)}
    \sig{\Big(\frac{L_y-y_\mb}{s}\Big)} \Bigr],
\end{multline}
where $(x_\mb, y_\mb)$ are the coordinates relative to one corner of the box. The box potential has the dimensions $L_x$ and $L_y$. The use of a sigmoid function 
\begin{equation}
    \sig(x) := \frac{1}{2} \left[1+\tanh\Big(\frac{x}{2}\Big)\right]
\end{equation} 
provides a smooth box boundary. Its width is controlled by the parameter $s$, which we set to $s=\SI{1.5}{\micro\meter}$. Finite values of $s$ improve the numerical integration by spreading the potential gradient over a non-zero width. The potential strength is set to $V_{\mb,0}/J = 100$.
The rotation of the box is implemented by applying a coordinate transformation between the box and lattice via 
\begin{equation}    
\begin{pmatrix}
    x\\
    y
\end{pmatrix}
=
\begin{pmatrix}
    \cos(\alpha) & -\sin(\alpha) \\
    \sin(\alpha) & \cos(\alpha)
\end{pmatrix}
\begin{pmatrix}
    x_\mb-L_x/2\\
    y_\mb-L_y/2
\end{pmatrix}
+
\begin{pmatrix}
    x_0 \\
    y_0
\end{pmatrix}
,
\end{equation}
where $\alpha$ is the angle of the rotating box 
and $(x_0, y_0)$ represents the center of the box. 
We linearly increase the angular velocity $\omega(t) := d\alpha/dt$ from $0$ to its final value over a duration of \SI{76}{\milli\second}. The final angular velocity is varied between $\omega = 0$ and $\SI{84.5}{\per\second}$.

\begin{figure}[h]
    \centering
    \includegraphics[]{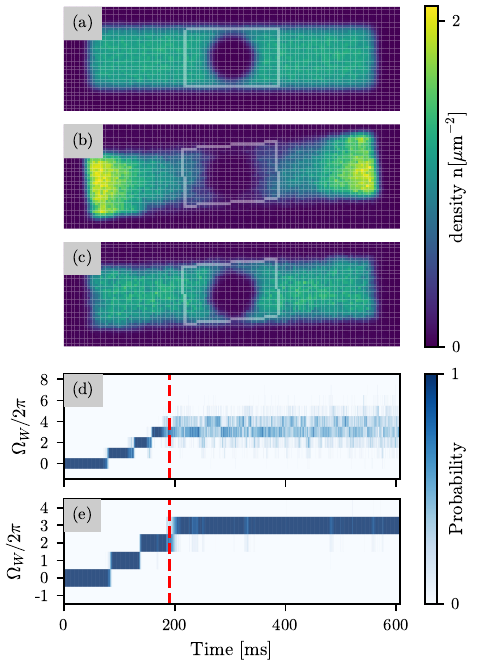}
    \caption{Condensate dynamics in a rotating box potential.
     (a) Initial density distribution $n(x,y)$ featuring a central depletion created by a repulsive Gaussian barrier, effectively dividing the condensate into two reservoirs connected by weak links  acting as Josephson junctions (JJs). 
     The bulk density is approximately \SI{1.2}{\per\micro\meter\squared}. 
     (b) Rotation at angular velocity $\omega=\SI{79}{\per\second}$ induces centrifugal acceleration, pushing atoms outward and enhancing central depletion.
     (c) The centrifugal effect is compensated by adding a harmonic confinement with frequency $\omega_h = \omega$, which restores the density profile and preserves reservoir connectivity. 
     In (a-c), a white rectangular contour indicates the path used to compute the global phase winding $\Omega_W$. 
     (d) Time evolution of the probability distribution of $\Omega_W$ without harmonic confinement. 
     The red line marks the time ($t= \SI{196}{ms}$) when the box reaches its final angular velocity. 
     Phase coherence across JJs is lost due to centrifugal depletion, leading to fluctuating values of $\Omega_W$. 
     (e) Same as (d), but with harmonic confinement applied. The phase winding stabilizes in the ground state. }
    \label{fig:cctest}
\end{figure}

To create the AQUID geometry, we use a Gaussian barrier potential of the form
\begin{equation}
V_{\mG} = V_{\mG,0}\exp\left({-\frac{(x-x_{0})^2+(y-y_{0})^2}{2\sigma^2}}\right),
\end{equation}
where $V_{\mG,0}$ is the strength and $\sigma$ is the width. 
We use $V_{\mG,0}/J = 5$ and several values for $\sigma$ in the range between \SI{3.5}{\micro\meter} and \SI{4}{\micro\meter}. 
This potential creates a density depletion region, effectively splitting the condensate into two reservoirs connected by weak links, see Fig. \ref{fig:cctest}. These weak links act as Josephson junctions (JJs) in our setup.

Due to the rotation of the box potential, the condensate experiences a centrifugal effect, which reduces the density in the JJs. This reduction disrupts the phase coherence between the two reservoirs, leading to strong fluctuations in phase winding [Fig. \ref{fig:cctest}]. 
This effect poses a bottleneck for implementing an atomtronic AQUID in the rotating box geometry. 
To mitigate this effect, we introduce an additional harmonic potential of the form
\begin{equation}
	V_{h} = \frac{1}{2}m\omega_h^2\left[(x-x_{0})^2+(y-y_{0})^2\right],
\end{equation}
where the trapping frequency $\omega_h$ is set equal to the angular velocity $\omega$. This choice ensures that the harmonic confinement exactly compensates the centrifugal acceleration experienced in the rotating frame, thereby stabilizing the density in the junction regions. 

Finally, we compute the observables, the density $n(\br, t)=|\psi(\br, t)|^2$ and the phase $\theta(\br, t)=\arg(\psi(\br, t))$, from the wavefunction $\psi(\br, t)$. To calculate the phase winding at each time step, we perform a numerical integration of the phase gradient $\Delta \theta$ along a rectangular path that encloses the central depletion zone and co-rotates with the box [Fig. \ref{fig:cctest}]. This path is discretized into $N=108$ equally spaced points, for each of which the nearest lattice site is selected.  The local phase difference $\delta \theta_k \in [-\pi,\pi)$ is then computed between neighboring points $k$ and $k+1$. The global phase winding $\Omega_W$ is obtained by summing over all local phase differences: $\Omega_W = \sum_{k=1}^{N} \delta \theta_k$. Both the density and the phase winding are averaged over the initial ensemble.

%
%

%
\section{Phase-slip dynamics}\label{section:phaseSlip}
We now characterize the operation of the AQUID by analyzing the phase-slip dynamics across different system parameters. Analogous to a toroidal AQUID, we expect the phase winding $\Omega_W$ to follow a step-like behavior as a function of angular velocity $\omega$.
Since the velocity of a condensate is proportional to its phase gradient, the single-valuedness of the wavefunction demands that the circulation, defined by the line integral of the velocity around a closed loop, is quantized \cite{pethick2008bose}.
In a 1D toroidal condensate, this condition directly implies the quantization of $\omega$, with an angular velocity quantum 
\begin{equation}\label{eq_ang_vel_quantum}
\omega_{\mt,0} = \frac{\hbar}{mr^2}
\end{equation}
where $r$ is the radius of the toroid \cite{pethick2008bose}.
When a barrier potential moves along the toroid at an arbitrary $\omega$, 
screening currents form in the condensate to maintain the quantization condition. 
If the barrier is sufficiently strong, the induced screening currents may exceed the critical current, 
making a phase slip energetically favorable. 
As a result, phase slips typically occur at half-integer multiples of $\omega_{\mt,0}$, 
where the sign of the screening current changes and its magnitude is minimized to $\le \omega_{\mt,0}/2$. 
This gives rise to a characteristic stepwise increase in the phase winding with increasing angular velocity of the rotating trap \cite{kiehn2022, mathey2016}.

In the case of a rotating box, however, the radius $r$ is not well-defined and the density around the Gaussian depletion is not uniform. Therefore, the equivalent of $\omega_{\mt,0}$ has no simple, geometric form. Nevertheless, we expect the relative velocity of the condensate with respect to the rotating potential to be minimized. Hence, phase slips should still occur near half-integer multiples of some effective angular velocity quantum $\omega_0$.
Due to the box's complex geometry, an analytical expression for $\omega_0$ is not straightforward, and we instead determine it numerically by fitting the phase-winding results to a step function.  

\begin{figure}
    \centering
    \includegraphics[]{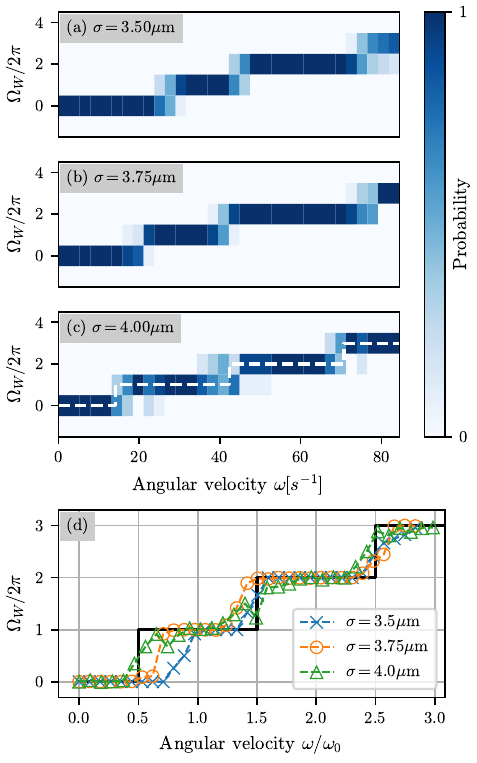}
    \caption{Phase winding statistics and angular velocity quantization.
    Probability distribution of phase winding $\Omega_W$, averaged over the time window $t=586-606\, \mathrm{ms}$, across $12$ samples for varying angular velocities $\omega$. 
    (a, b) For barrier widths $\sigma= 3.5$ and $3.75\, \mu \mathrm{m}$, the phase-winding plateaus exhibit unequal widths, indicating non-ideal AQUID behavior.  
    (c) For $\sigma = \SI{4}{\micro\meter}$, the winding states occur with nearly equal width, 
    consistent with ideal AQUID operation. The fit to a step function (white dashed line) yields an  angular velocity quantum $\omega_0= \SI{28.3 \pm 0.3}{\per\s}$. 
    (d) Ensemble-averaged phase winding as a function of $\omega/\omega_0$ for all three cases. The solid line represents the ideal AQUID step-function prediction. 
    }
    \label{fig:phasewinding_different_sigmas}
\end{figure}

To this end, we plot the probability of each phase winding number, averaged over the final \SI{20}{ms} of the simulation time, as a function of the angular velocity $\omega$, see Fig. \ref{fig:phasewinding_different_sigmas}. 
This is done for three different barrier widths $\sigma$. 
In all cases, the phase winding approximately follows a step function, as expected.
For $\sigma = \SI{3.5}{\micro\meter}$ and $\SI{3.75}{\micro\meter}$, the step spacings are unequal. 
However, for $\sigma = \SI{4}{\micro\meter}$, the steps are nearly equal in spacing, 
consistent with an ideal AQUID behavior. 
The unequal step spacings at lower barrier widths are attributed to the onset of phase slips occurring at higher $\omega$, particularly evident in the first step.
This is likely due to the dependence of the critical velocity on the barrier width: 
for narrower barriers, the coupling between the reservoirs is stronger, increasing the critical velocity \cite{Singh2020jj}. 
As a result, larger screening currents are needed to trigger a phase slip, causing the slips to occur at 
higher $\omega$ than expected. This deviation from half-integer quantization reduces the quality of AQUID operation. Similar deviations have been observed in toroidal condensates \cite{mathey2016, kiehn2022}.

We also find that the variance of the phase winding increases considerably with barrier width. 
This can be attributed to the reduced coupling between the reservoirs at larger $\sigma$, 
which suppresses phase coherence and increases the noise of the phase winding numbers. 
While $\sigma = \SI{4}{\micro\meter}$ provides near-ideal AQUID behavior in terms of step uniformity, the associated increase in fluctuations suggests a trade-off. 
At even higher values of  $\sigma$, the noise may become large enough to inhibit reliable AQUID operation. We therefore identify $\sigma = \SI{4}{\micro\meter}$ as near-optimal choice for our system.

To determine the angular velocity quantum of the system, we fit a step function to the phase-winding results at $\sigma = \SI{4}{\micro\meter}$ in Fig. \ref{fig:phasewinding_different_sigmas}(c), yielding a value of $\omega_0 = \SI{28.3 \pm 0.3}{\per\s}$. This is significantly smaller than the value obtained using Eq. \ref{eq_ang_vel_quantum} together with the distance of the Josephson junctions from the rotation axis, $r_{\rm{JJ}} \approx \SI{9}{\micro\meter}$, which gives $\omega_{\mt,0} = \SI{65.2}{\per\second}$. 
This discrepancy highlights the importance of accounting for the full bulk of the condensate in the outer regions of the box potential when determining $\omega_0$. 
Because the outer regions have higher linear velocities, they facilitate phase slips at lower $\omega$ compared to a  toroidal condensate lacking large reservoirs.
Based on Eq. \ref{eq_ang_vel_quantum} and the extracted $\omega_0$, 
we obtain an effective radius of $r_{\rm{eff}} \approx \SI{13.3}{\micro\meter}$, which is considerably larger than $r_{\rm{JJ}}$ defined solely by the junction distance from the rotation axis. 
 %




\begin{figure}
    \centering
    \includegraphics[]{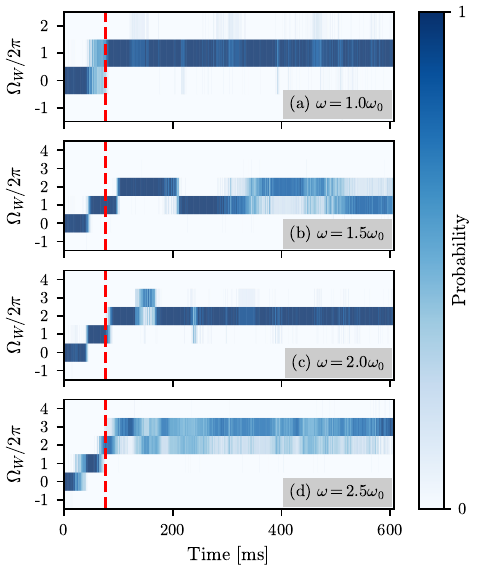}
    \caption{
    Time evolution of phase-winding probability for optimal barrier width $\sigma =\SI{4}{\micro\meter}$.
    (a, c) At angular velocities $\omega/\omega_0 = 1,\, 2$, corresponding to integer flux quanta, 
    the system quickly relaxes to well defined phase-winding states shortly after the rotation ramp is completed.  
    (b, d) For $\omega/\omega_0 = 1.5,\, 2.5$, i.e., near half-integer flux quanta, the system enters a metastable regime characterized by oscillations between adjacent phase-winding states. 
    The vertical dashed red line marks the acceleration time over which $\omega$ is linearly ramped up to its final value. }
    \label{fig:phasewinding_time_development}
\end{figure}

To investigate how the AQUID's relaxation depends on angular velocity, we analyze the time evolution of the phase winding at fixed final angular velocities $\omega$ in Fig. \ref{fig:phasewinding_time_development}. 
The box reaches its final angular velocity at $t=\SI{76}{ms}$. 
When $\omega$ is close to an integer multiple of $\omega_0$ [Figs. \ref{fig:phasewinding_time_development}(a, c)], 
the AQUID quickly relaxes to the ground state within \SI{200}{ms}. 
In contrast, when $\omega$ is close to a step, i.e., close to half-integer multiples of $\omega_0$, 
the system fails to relax even within \SI{600}{ms} simulation time [Figs. \ref{fig:phasewinding_time_development}(b, d)]. 
Instead, it remains in a metastable state,  oscillating between two adjacent phase-winding values, as seen in Fig. \ref{fig:phasewinding_time_development}(b). Similar metastable oscillations have been previously discussed in toroidal condensates \cite{mathey2016}.

This variation in relaxation behavior can be understood in terms of the energy difference between phase winding states. 
When $\omega$ is near an integer multiple of $\omega_0$, i.e., $\omega= n \omega_0$ $(n=0,\, 1,\,2,\, ...)$, 
screening currents vanish in the ground state and are large if the phase winding does not match the box's rotation. This results in a sizable energy gap favoring relaxation to winding number of lowest energy. 
However, near half-integer multiples, $\omega= (n+1/2)\omega_0$, the screening currents for phase windings $n$ and $n+1$ are equal in magnitude but opposite in sign. Thus, the ground state becomes degenerate, reducing the energy benefit of a phase slip and suppressing  relaxation \cite{kiehn2022}. This also explains the increased variance observed near the step transitions in Fig.  \ref{fig:phasewinding_different_sigmas}.

\begin{figure}
    \includegraphics[]{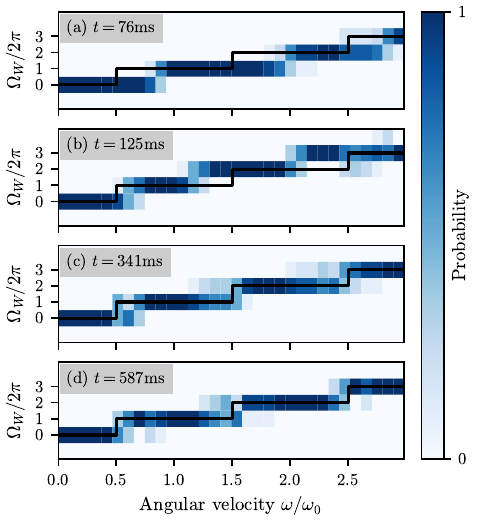}
    \caption{Phase-winding probability as a function of $\omega/\omega_0$, shown at successive  times (averaged over a $\SI{20}{ms}$ window starting at time $t$). 
    (a) At $t=\SI{76}{ms}$, the box reaches its final angular velocity, but the system has not yet relaxed to the ground state. 
    (b) At $t=\SI{125}{ms}$, some samples are excited to phase-winding states higher than the energetically favored state. 
    (c) At $t=\SI{341}{ms}$, the AQUID has largely relaxed to the ground state, except near the half-flux quantum. 
    (d) Continued relaxation further improves the steps near half-flux quantum values. 
    The solid line shows the ideal AQUID step-function prediction.}
    \label{fig:phasewinding_different_times}
\end{figure}

The angular velocity dependence of relaxation time is further illustrated by examining the temporal evolution of the step structure [Fig. \ref{fig:phasewinding_different_times}]. 
At $t=\SI{76}{ms}$, when the box first reaches its final angular velocity, steps appear at approximately $\omega_0$ and $2\omega_0$, rather than at half-integer multiples. This delay occurs because the system requires additional time to relax into the ground state after reaching the critical angular velocity $(n+1/2)\omega_0$. 
At higher angular velocities, the larger acceleration causes this threshold to be reached earlier, 
so the system has sufficient time to relax. 
By $t=\SI{125}{ms}$, the step edges have shifted toward lower $\omega$. 
Interestingly, some realizations in the regions between $\omega_0$ and $1.5\omega_0$, and between $2\omega_0$ and $2.5\omega_0$, are excited to higher phase winding states than the energetically favored value. 
By $t=\SI{341}{ms}$, the AQUID has relaxed into the ground state for most $\omega$, except near $(n+1/2)\omega_0$, 
where relaxation remains suppressed due to the near-degeneracy of phase-winding states. 
Over the subsequent relaxation, the step structure sharpens only marginally, indicating that the relaxation dynamics are fundamentally limited near these degeneracy points. 




\section{Underlying mechanism}\label{sec:time_phase_slip}

\begin{figure}
\includegraphics[width=0.9\linewidth]{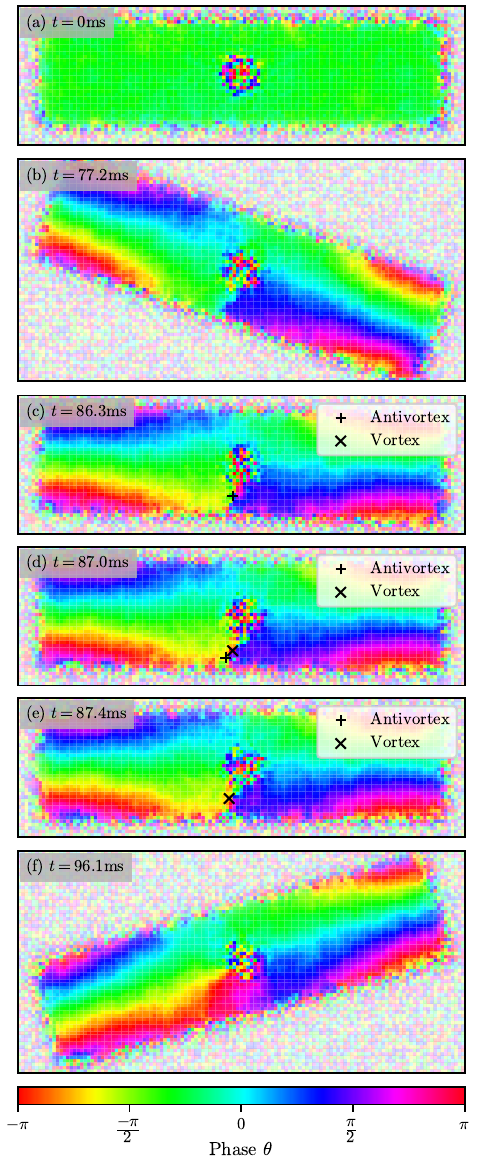}
    \caption{Time evolution of the phase distribution $\theta(x, y)$ for a single sample. 
    (a) At $t=0$, the phase variation within the bulk of the condensate is minimal. 
    (b) As the box accelerates, screening currents develop across the JJs, with the bottom JJ exhibiting a larger phase difference than the top. (c) An antivortex (indicated by plus) enters from the central depletion zone. (d) While the antivortex moves toward the edge, leading to an increase of the global phase winding by $2\pi$, a vortex (indicated by cross) enters from the central depletion zone. 
    (e) The antivortex is eventually absorbed at the edge of the cloud 
    and the vortex returns to the central depletion zone and is absorbed there. 
    (f) At later times, the phase gradient spreads over a broader region of the condensate.}
    \label{fig:phase_2d}
\end{figure}

To illustrate how a non-zero phase winding is generated, we show the phase distribution $\theta(x,y)$ of a single sample at various times $t$ in Fig. \ref{fig:phase_2d}, 
    for a final angular velocity of $\omega = \SI{32}{\per\second}$, reached at $t=\SI{76}{ms}$. 
For visual interpretation, fast phase oscillations have been removed by multiplying the wavefunction by a factor $\exp(-i\beta t/\hbar)$, with $\beta \approx \SI{8.9}{J}$.
Vortices are identified by computing the net phase winding around a lattice plaquette of size $l\times l$ using $\sum_{\Box} \delta \theta(x,y) = \delta_x \theta(x,y) + \delta_y\theta(x+l,y)+\delta_x\theta(x+l,y+l)+\delta_y\theta(x,y+l)$, 
where the phase differences between sites are taken to be $\delta_{x/y} \theta(x,y)  \in (-\pi, \pi]$. 
A net phase winding of $2\pi$ ($-2\pi$) indicates a vortex (antivortex).
Vortices located inside the central depletion zone and outside the box are omitted.

\begin{figure}
    \includegraphics[]{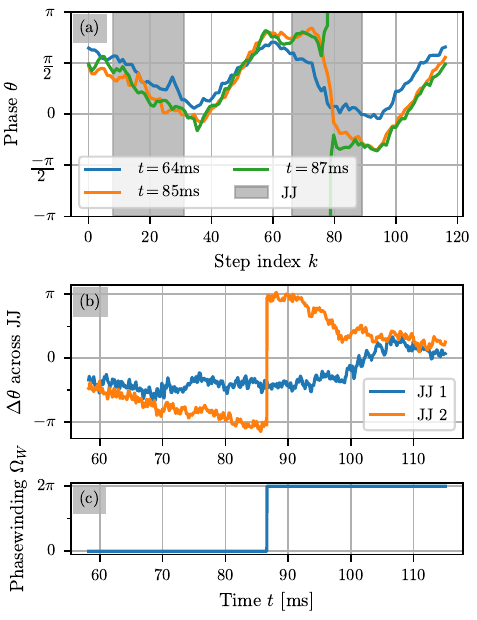}
    \caption{
     (a) Phase evolution along the integration path (denoted by discrete path index $k$). 
     For $t<\SI{87}{ms}$, the phase gradient is negative within the JJs, indicating clockwise screening currents. Outside the JJs, the gradient is positive, showing that the condensate co-rotates counter-clockwise 
     with the box. 
     (b) Phase difference $\Delta \theta (t)$ across the Josephson junctions (JJs). 
     In the second junction (JJ2), the phase gradient gradually steepens until $\Delta \theta$ approaches nearly $-\pi$ at $t \approx \SI{86}{ms}$, at which point a phase slip occurs. 
     This event changes the sign of $\Delta \theta$, thus reversing the screening current direction. 
     Subsequently, $\Delta \theta$ across JJ2 decreases, 
     while $\Delta \theta$ across JJ1 slowly increases. Eventually, both junctions stabilize at similar positive phase differences, indicating counter-clockwise screening currents in both JJs. 
    (c) Time evolution of the total phase winding.}
    \label{fig:phase_JJ}
\end{figure}

At $t=0$, the phase inside the box in the bulk is nearly spatially uniform, see Fig. \ref{fig:phase_2d}(a). In contrast, both the central depletion zone and outside of the box exhibit strong phase fluctuations. 
This implies that a sufficiently smooth phase gradient is only well-defined on a multiply connected region, a prerequisite for stable phase winding around the central barrier.
By $t=\SI{77.2}{ms}$ [Fig. \ref{fig:phase_2d}(b)], the phase gradient in the outer condensate points counterclockwise and becomes steeper with increasing radial distance, consistent  with the rigid-body-like rotation of the box potential. 
The flow remains irrotational, as required for a superfluid. 
Within the Josephson junctions (JJs), however, the phase gradient points clockwise, indicating the formation of screening currents and a net phase winding of zero around the center 
[Fig. \ref{fig:phase_JJ}(a)]. 

As the box continues to accelerate, the phase difference across one of the JJs increases, eventually reaching approximately $-\pi$, while the other remains relatively unchanged, breaking the symmetry [Fig. \ref{fig:phase_JJ}(b)]. 
The phase on either side of this junction approaches $-\pi/2$ (green) and $\pi/2$ (blue).
 At $t = \SI{86.3}{ms}$ [Fig. \ref{fig:phase_2d}(c)], an antivortex enters the JJ from the central depletion zone. 
 This is visually evident from phase values near $+\pi$ (red) intruding into the junction region. 
 As the antivortex travels toward the condensate edge, it inverts the local screening current direction between itself and the depletion zone, thereby increasing the global phase winding to $2\pi$ for all integration paths passing through that region 
 [Fig. \ref{fig:phase_JJ}(c)].
Shortly after, at $t = \SI{87}{ms}$ [Fig. \ref{fig:phase_2d}(d)], a vortex enters from the central depletion zone. 
This temporarily reduces the global phase winding to zero for integration paths located between the vortex and the central depletion zone. 
The antivortex continues moving outward and is eventually absorbed at the condensate boundary.   

Subsequently, the vortex is reabsorbed in the central depletion zone, restoring a global phase winding of $2\pi$ across all integration paths [Fig. \ref{fig:phase_2d}(e)]. 
As the system relaxes, the phase difference across the original JJ decreases and spreads over a wider area [Fig. \ref{fig:phase_2d}(f)]. Meanwhile, the phase difference across the other JJ increases until both junctions reach similar positive values, indicating counter-clockwise screening currents in both after the phase slip [Fig. \ref{fig:phase_JJ}(b)].

A video illustrating these vortex dynamics is included in the supplementary material. 
Phase-slip dynamics driven by a single antivortex, as observed here, have also been reported in toroidal condensates \cite{Piazza2009, Mathey2014}.
Since only a single realization is presented, we note that the specific mechanism of the phase slip may vary across different  samples. In some cases, the phase slip may instead be mediated by a vortex propagating from the edge toward the center, 
rather than by an antivortex moving outward, as seen in our example.


\section{Voltage-flux relation}\label{sec:voltage_state}
A characteristic property of any SQUID is the periodic dependence of voltage on magnetic flux \cite{gross2016}. In an atomtronic AQUID, the analog of voltage is the chemical-potential difference between the reservoirs, which is directly proportional to the density imbalance \cite{kiehn2022}. 
These voltage oscillations originate from interference between the currents through the two Josephson junctions, which results in a modulation of the critical current as a function of the applied flux \cite{gross2016}. Quantitatively, this can be understood in terms of the screening currents.
At half-integer multiples of the flux quantum $\Phi_0$, the screening currents are large. As a result, 
the total current (bias plus screening) exceeds the critical current of one of the junctions by a significant value, driving the SQUID deep into the resistive regime, and resulting in a high voltage. 
In contrast, at integer multiples of $\Phi_0$, the screening currents are minimal, leading to a low voltage \cite{clarke2006squid}. Therefore, we expect the density imbalance to exhibit the same periodicity with angular velocity $\omega$ as the screening current,  showing maxima at half-integer multiples of $\omega_0$ and minima at integer multiples \cite{kiehn2022}. 
In the limit of negligible screening and strong damping, the current-voltage characteristic of a SQUID is described by \cite{gross2016}
\begin{equation}\label{eq:IVC_negligible_screening}
    V=I_{c} R\sqrt{\left(\frac{I}{2 I_{c}}\right)^{2}-\left[\cos \left(\pi \frac{\Phi}{\Phi_{0}}\right)\right]^{2}},
\end{equation}
where $I$ is the bias current, $I_c$ is the critical current of a single junction, $\Phi$ is the flux, and $R$ is the resistance.

\begin{figure}
    \includegraphics{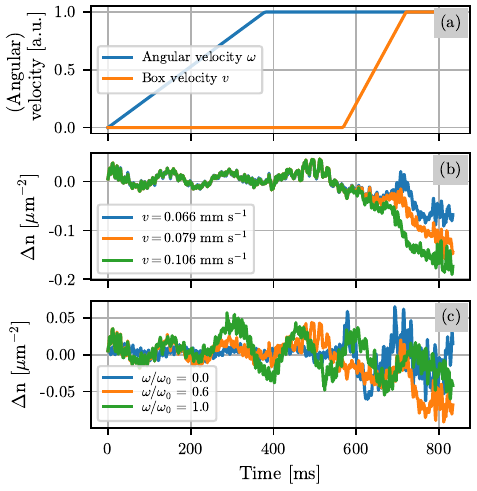}
    \caption{
    (a) Protocol for creating a density imbalance in the rotating box geometry. 
    The angular velocity $\omega$ is ramped up linearly over \SI{379}{ms}, followed by a constant rotation time of \SI{190}{ms}. Subsequently, the box velocity $v$ is ramped up linearly to its final value over \SI{152}{ms}.
     Simulation results showing the dynamics of the density imbalance $\Delta n$ between the reservoirs: 
     (b) For varying $v$ at fixed $\omega$; 
     (c) For varying $\omega$ at fixed $v$. 
    }
    \label{fig:density_imbalance_time_evolution}
\end{figure}

To probe this behavior, we introduce a density imbalance by moving the box potential at velocity $v$ along its long axis, 
while keeping the Gaussian barrier fixed in position. 
The protocol is as follows. We linearly ramp up $\omega$ over \SI{379}{ms}, let the box rotate for a duration of \SI{190}{ms}, 
and then ramp up the box velocity $v$ over \SI{152}{ms}, 
as illustrated in Fig. \ref{fig:density_imbalance_time_evolution}(a). 
We continue the simulation for additional \SI{265}{ms}, which results in a sizable density imbalance  $\Delta n$ between the reservoirs, 
determined by averaging the density in both reservoirs and calculating their difference.
Figs. \ref{fig:density_imbalance_time_evolution}(b,c) show the time evolution $\Delta n(t)$ for selected combinations of $\omega$ and $v$.
From the start of the simulation, we observe oscillations with a frequency of about \SI{7}{Hz}, which decreases to about \SI{4}{Hz} over time as $\omega$ increases. 
These are attributed to plasma oscillations, consistent with previous studies \cite{Albiez2005, Levy2007, kiehn2022}. 
For constant $\omega$, increasing $v$ leads to stronger bias currents and therefore a larger final imbalance. 
For constant $v = 0.066\, \mms$, the final imbalance is highest near $\omega \approx 0.5 \omega_0$, 
while at $\omega=0$ and $\omega =\omega_0$, the imbalance is reduced, consistent with the expected periodicity.

\begin{figure}
    \includegraphics[]{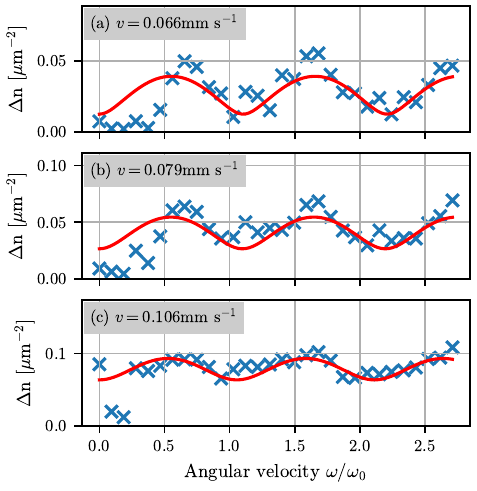}
    \caption{
    Imbalance-flux characteristics. 
    Time-averaged density imbalance $\Delta n$ as a function of angular velocity $\omega$, along with the fits to Eq. \ref{eq:IVC_negligible_screening} (continuous lines). 
    Results are shown for three box velocities (a) $v = 0.066$, (b) $0.079$, and (c) $0.106\, \mms$.
    }
    \label{fig:density_imbalance}
\end{figure}

To suppress the influence of plasma oscillations, we compute the time-averaged imbalance $\Delta n$
over the last \SI{265}{ms} of the simulation. 
In Fig. \ref{fig:density_imbalance}, we show the averaged $\Delta n$ as a function of $\omega$, for three different box velocities $v = 0.066,\ 0.079$ and $0.106\, \mms$.
For all three $v$, the expected periodic modulation of $\Delta n (\omega)$ is clearly visible.
While $\Delta n$ increases with $v$, the strength of modulation decreases. 
Both observations are in accordance with Eq. \ref{eq:IVC_negligible_screening}, 
which predicts $\Delta n$ to increase with $v$, and modulation magnitude to decrease as the bias current term $I/(2I_c)$ dominates over the oscillation term. 
From the fits to Eq. \ref{eq:IVC_negligible_screening}, we extract the normalized bias currents $I/(2I_c) = 1.03 \pm 0.05$, $1.12 \pm 0.08$, and $1.37 \pm 0.16$, for box velocities $v = 0.066,\ 0.079$ and $0.106\, \mms$, respectively. 
These values indicate that at low $v$, the system operates near the transition to the resistive regime, 
while higher $v$ pushes the system deeper into the resistive regime. 
This confirms that $v = 0.066\,\mms$ offers optimal conditions for dc-AQUID operation, 
providing clear density imbalance modulation with minimal dissipation.

The period of modulation agrees well with the previously extracted angular velocity quantum $\omega_0 = \SI{28.3 \pm 0.3}{\per\s}$, 
obtained from the phase-winding step analysis in Sec. \ref{section:phaseSlip}.  
The fits to Eq. \ref{eq:IVC_negligible_screening} yield $\omega_0= \SI{31.2 \pm 0.9}{}$, $\omega_0= \SI{31.0 \pm 0.9}{}$, and $\omega_0= \SI{29.8 \pm 1.0}{\per\s}$, respectively, for increasing $v$. 
These values are systematically higher than the step-function estimate by about $1$ to \SI{2}{Hz}, 
possibly due to nonequilibrium effects or finite-size corrections. 
Interestingly, across all simulations, the imbalance between $0$ and $\SI{0.5}{\omega_0}$ is lower than expected, 
suggesting a possibly enhanced critical current in this regime. This may be due to weaker plasma oscillations, 
and aligns with delayed onset of the first phase-winding step observed in Fig. \ref{fig:phasewinding_different_sigmas}.



\section{Conclusions} \label{sec:con}
We have presented an alternative configuration for implementing an atomtronic AQUID, based on a BEC confined in a rotating box potential with a central repulsive barrier. This barrier effectively splits the condensate into two reservoirs connected by two weak links, acting as Josephson junctions (JJs). The rotation of the box induces a centrifugal depletion of the condensate, 
which suppresses phase coherence across the JJs and poses a challenge for AQUID operation. 
We show that this effect can be mitigated by introducing a counteracting harmonic potential with a trapping frequency matched to the angular velocity of the box. 
By tuning the central barrier width, we optimize the phase-winding characteristics, achieving an ideal step-like dependence of the global phase winding on the angular velocity. Overall, the system relaxes to the ground state within approximately \SI{200}{ms}. 
Our simulations further reveal that phase slips are mediated by an antivortex propagating  from the central depletion region to the edge of the condensate through one of the JJs---a microscopic mechanism for the quantized change in circulation.
To probe the voltage-like response of the AQUID, we translate the box potential laterally relative to the fixed central barrier, 
inducing a density imbalance between the reservoirs. This imbalance exhibits a clear periodic dependence on angular velocity, consistent with the expected voltage-flux characteristics of a SQUID.

These findings demonstrate that rotating box potentials provide a viable and versatile alternative to toroidal 
geometries for realizing atomtronic AQUIDs. This platform retains key features such as quantized circulation and interference, 
while offering enhanced flexibility in junction placement and dynamic trap control. 
As such, it opens new opportunities for investigating quantum transport, rotating sensing, and non-equilibrium phase dynamics in superfluid systems.

	\section*{Acknowledgements} 
    L.M. acknowledges support by the Deutsche Forschungsgemeinschaft (DFG, German Research Foundation), namely the Cluster of Excellence ‘Advanced Imaging of Matter’ (EXC 2056), Project No. 390715994. 
    The project is co-financed by ERDF of the European Union and by ‘Fonds of the Hamburg Ministry of Science, Research, Equalities and Districts (BWFGB)’.
    



%
%
\appendix
\setcounter{equation}{0}
\setcounter{figure}{0}
\setcounter{table}{0}
\setcounter{section}{0}
\setcounter{subsection}{0}
\renewcommand{\thefigure}{A\arabic{figure}}
\renewcommand{\thetable}{A\arabic{table}}
%

%
\bibliography{BA2}

\providecommand{\noopsort}[1]{}\providecommand{\singleletter}[1]{#1}%
\begin{thebibliography}{46}%
\makeatletter
\providecommand \@ifxundefined [1]{%
 \@ifx{#1\undefined}
}%
\providecommand \@ifnum [1]{%
 \ifnum #1\expandafter \@firstoftwo
 \else \expandafter \@secondoftwo
 \fi
}%
\providecommand \@ifx [1]{%
 \ifx #1\expandafter \@firstoftwo
 \else \expandafter \@secondoftwo
 \fi
}%
\providecommand \natexlab [1]{#1}%
\providecommand \enquote  [1]{``#1''}%
\providecommand \bibnamefont  [1]{#1}%
\providecommand \bibfnamefont [1]{#1}%
\providecommand \citenamefont [1]{#1}%
\providecommand \href@noop [0]{\@secondoftwo}%
\providecommand \href [0]{\begingroup \@sanitize@url \@href}%
\providecommand \@href[1]{\@@startlink{#1}\@@href}%
\providecommand \@@href[1]{\endgroup#1\@@endlink}%
\providecommand \@sanitize@url [0]{\catcode `\\12\catcode `\$12\catcode `\&12\catcode `\#12\catcode `\^12\catcode `\_12\catcode `\%12\relax}%
\providecommand \@@startlink[1]{}%
\providecommand \@@endlink[0]{}%
\providecommand \url  [0]{\begingroup\@sanitize@url \@url }%
\providecommand \@url [1]{\endgroup\@href {#1}{\urlprefix }}%
\providecommand \urlprefix  [0]{URL }%
\providecommand \Eprint [0]{\href }%
\providecommand \doibase [0]{https://doi.org/}%
\providecommand \selectlanguage [0]{\@gobble}%
\providecommand \bibinfo  [0]{\@secondoftwo}%
\providecommand \bibfield  [0]{\@secondoftwo}%
\providecommand \translation [1]{[#1]}%
\providecommand \BibitemOpen [0]{}%
\providecommand \bibitemStop [0]{}%
\providecommand \bibitemNoStop [0]{.\EOS\space}%
\providecommand \EOS [0]{\spacefactor3000\relax}%
\providecommand \BibitemShut  [1]{\csname bibitem#1\endcsname}%
\let\auto@bib@innerbib\@empty
\bibitem [{\citenamefont {Amico}\ \emph {et~al.}(2021)\citenamefont {Amico}, \citenamefont {Boshier}, \citenamefont {Birkl}, \citenamefont {Minguzzi}, \citenamefont {Miniatura}, \citenamefont {Kwek}, \citenamefont {Aghamalyan}, \citenamefont {Ahufinger}, \citenamefont {Anderson}, \citenamefont {Andrei}, \citenamefont {Arnold}, \citenamefont {Baker}, \citenamefont {Bell}, \citenamefont {Bland}, \citenamefont {Brantut}, \citenamefont {Cassettari}, \citenamefont {Chetcuti}, \citenamefont {Chevy}, \citenamefont {Citro}, \citenamefont {De~Palo}, \citenamefont {Dumke}, \citenamefont {Edwards}, \citenamefont {Folman}, \citenamefont {Fortagh}, \citenamefont {Gardiner}, \citenamefont {Garraway}, \citenamefont {Gauthier}, \citenamefont {Günther}, \citenamefont {Haug}, \citenamefont {Hufnagel}, \citenamefont {Keil}, \citenamefont {Ireland}, \citenamefont {Lebrat}, \citenamefont {Li}, \citenamefont {Longchambon}, \citenamefont {Mompart}, \citenamefont {Morsch}, \citenamefont {Naldesi}, \citenamefont {Neely},
  \citenamefont {Olshanii}, \citenamefont {Orignac}, \citenamefont {Pandey}, \citenamefont {Pérez-Obiol}, \citenamefont {Perrin}, \citenamefont {Piroli}, \citenamefont {Polo}, \citenamefont {Pritchard}, \citenamefont {Proukakis}, \citenamefont {Rylands}, \citenamefont {Rubinsztein-Dunlop}, \citenamefont {Scazza}, \citenamefont {Stringari}, \citenamefont {Tosto}, \citenamefont {Trombettoni}, \citenamefont {Victorin}, \citenamefont {Klitzing}, \citenamefont {Wilkowski}, \citenamefont {Xhani},\ and\ \citenamefont {Yakimenko}}]{Amico2021}%
  \BibitemOpen
  \bibfield  {author} {\bibinfo {author} {\bibfnamefont {L.}~\bibnamefont {Amico}}, \bibinfo {author} {\bibfnamefont {M.}~\bibnamefont {Boshier}}, \bibinfo {author} {\bibfnamefont {G.}~\bibnamefont {Birkl}}, \bibinfo {author} {\bibfnamefont {A.}~\bibnamefont {Minguzzi}}, \bibinfo {author} {\bibfnamefont {C.}~\bibnamefont {Miniatura}}, \bibinfo {author} {\bibfnamefont {L.-C.}\ \bibnamefont {Kwek}}, \bibinfo {author} {\bibfnamefont {D.}~\bibnamefont {Aghamalyan}}, \bibinfo {author} {\bibfnamefont {V.}~\bibnamefont {Ahufinger}}, \bibinfo {author} {\bibfnamefont {D.}~\bibnamefont {Anderson}}, \bibinfo {author} {\bibfnamefont {N.}~\bibnamefont {Andrei}}, \bibinfo {author} {\bibfnamefont {A.~S.}\ \bibnamefont {Arnold}}, \bibinfo {author} {\bibfnamefont {M.}~\bibnamefont {Baker}}, \bibinfo {author} {\bibfnamefont {T.~A.}\ \bibnamefont {Bell}}, \bibinfo {author} {\bibfnamefont {T.}~\bibnamefont {Bland}}, \bibinfo {author} {\bibfnamefont {J.~P.}\ \bibnamefont {Brantut}}, \bibinfo {author} {\bibfnamefont
  {D.}~\bibnamefont {Cassettari}}, \bibinfo {author} {\bibfnamefont {W.~J.}\ \bibnamefont {Chetcuti}}, \bibinfo {author} {\bibfnamefont {F.}~\bibnamefont {Chevy}}, \bibinfo {author} {\bibfnamefont {R.}~\bibnamefont {Citro}}, \bibinfo {author} {\bibfnamefont {S.}~\bibnamefont {De~Palo}}, \bibinfo {author} {\bibfnamefont {R.}~\bibnamefont {Dumke}}, \bibinfo {author} {\bibfnamefont {M.}~\bibnamefont {Edwards}}, \bibinfo {author} {\bibfnamefont {R.}~\bibnamefont {Folman}}, \bibinfo {author} {\bibfnamefont {J.}~\bibnamefont {Fortagh}}, \bibinfo {author} {\bibfnamefont {S.~A.}\ \bibnamefont {Gardiner}}, \bibinfo {author} {\bibfnamefont {B.~M.}\ \bibnamefont {Garraway}}, \bibinfo {author} {\bibfnamefont {G.}~\bibnamefont {Gauthier}}, \bibinfo {author} {\bibfnamefont {A.}~\bibnamefont {Günther}}, \bibinfo {author} {\bibfnamefont {T.}~\bibnamefont {Haug}}, \bibinfo {author} {\bibfnamefont {C.}~\bibnamefont {Hufnagel}}, \bibinfo {author} {\bibfnamefont {M.}~\bibnamefont {Keil}}, \bibinfo {author} {\bibfnamefont
  {P.}~\bibnamefont {Ireland}}, \bibinfo {author} {\bibfnamefont {M.}~\bibnamefont {Lebrat}}, \bibinfo {author} {\bibfnamefont {W.}~\bibnamefont {Li}}, \bibinfo {author} {\bibfnamefont {L.}~\bibnamefont {Longchambon}}, \bibinfo {author} {\bibfnamefont {J.}~\bibnamefont {Mompart}}, \bibinfo {author} {\bibfnamefont {O.}~\bibnamefont {Morsch}}, \bibinfo {author} {\bibfnamefont {P.}~\bibnamefont {Naldesi}}, \bibinfo {author} {\bibfnamefont {T.~W.}\ \bibnamefont {Neely}}, \bibinfo {author} {\bibfnamefont {M.}~\bibnamefont {Olshanii}}, \bibinfo {author} {\bibfnamefont {E.}~\bibnamefont {Orignac}}, \bibinfo {author} {\bibfnamefont {S.}~\bibnamefont {Pandey}}, \bibinfo {author} {\bibfnamefont {A.}~\bibnamefont {Pérez-Obiol}}, \bibinfo {author} {\bibfnamefont {H.}~\bibnamefont {Perrin}}, \bibinfo {author} {\bibfnamefont {L.}~\bibnamefont {Piroli}}, \bibinfo {author} {\bibfnamefont {J.}~\bibnamefont {Polo}}, \bibinfo {author} {\bibfnamefont {A.~L.}\ \bibnamefont {Pritchard}}, \bibinfo {author} {\bibfnamefont {N.~P.}\
  \bibnamefont {Proukakis}}, \bibinfo {author} {\bibfnamefont {C.}~\bibnamefont {Rylands}}, \bibinfo {author} {\bibfnamefont {H.}~\bibnamefont {Rubinsztein-Dunlop}}, \bibinfo {author} {\bibfnamefont {F.}~\bibnamefont {Scazza}}, \bibinfo {author} {\bibfnamefont {S.}~\bibnamefont {Stringari}}, \bibinfo {author} {\bibfnamefont {F.}~\bibnamefont {Tosto}}, \bibinfo {author} {\bibfnamefont {A.}~\bibnamefont {Trombettoni}}, \bibinfo {author} {\bibfnamefont {N.}~\bibnamefont {Victorin}}, \bibinfo {author} {\bibfnamefont {W.~v.}\ \bibnamefont {Klitzing}}, \bibinfo {author} {\bibfnamefont {D.}~\bibnamefont {Wilkowski}}, \bibinfo {author} {\bibfnamefont {K.}~\bibnamefont {Xhani}},\ and\ \bibinfo {author} {\bibfnamefont {A.}~\bibnamefont {Yakimenko}},\ }\bibfield  {title} {\bibinfo {title} {{Roadmap on Atomtronics: State of the art and perspective}},\ }\bibfield  {journal} {\bibinfo  {journal} {AVS Quantum Science}\ }\textbf {\bibinfo {volume} {3}},\ \href {https://doi.org/10.1116/5.0026178} {10.1116/5.0026178} (\bibinfo
  {year} {2021}),\ \bibinfo {note} {039201}\BibitemShut {NoStop}%
\bibitem [{\citenamefont {Amico}\ \emph {et~al.}(2022)\citenamefont {Amico}, \citenamefont {Anderson}, \citenamefont {Boshier}, \citenamefont {Brantut}, \citenamefont {Kwek}, \citenamefont {Minguzzi},\ and\ \citenamefont {von Klitzing}}]{Amico2022}%
  \BibitemOpen
  \bibfield  {author} {\bibinfo {author} {\bibfnamefont {L.}~\bibnamefont {Amico}}, \bibinfo {author} {\bibfnamefont {D.}~\bibnamefont {Anderson}}, \bibinfo {author} {\bibfnamefont {M.}~\bibnamefont {Boshier}}, \bibinfo {author} {\bibfnamefont {J.-P.}\ \bibnamefont {Brantut}}, \bibinfo {author} {\bibfnamefont {L.-C.}\ \bibnamefont {Kwek}}, \bibinfo {author} {\bibfnamefont {A.}~\bibnamefont {Minguzzi}},\ and\ \bibinfo {author} {\bibfnamefont {W.}~\bibnamefont {von Klitzing}},\ }\bibfield  {title} {\bibinfo {title} {{Colloquium: Atomtronic circuits: From many-body physics to quantum technologies}},\ }\href {https://doi.org/10.1103/RevModPhys.94.041001} {\bibfield  {journal} {\bibinfo  {journal} {Rev. Mod. Phys.}\ }\textbf {\bibinfo {volume} {94}},\ \bibinfo {pages} {041001} (\bibinfo {year} {2022})}\BibitemShut {NoStop}%
\bibitem [{\citenamefont {Seaman}\ \emph {et~al.}(2007)\citenamefont {Seaman}, \citenamefont {Kr\"amer}, \citenamefont {Anderson},\ and\ \citenamefont {Holland}}]{AtomtronicsFirstPaper}%
  \BibitemOpen
  \bibfield  {author} {\bibinfo {author} {\bibfnamefont {B.~T.}\ \bibnamefont {Seaman}}, \bibinfo {author} {\bibfnamefont {M.}~\bibnamefont {Kr\"amer}}, \bibinfo {author} {\bibfnamefont {D.~Z.}\ \bibnamefont {Anderson}},\ and\ \bibinfo {author} {\bibfnamefont {M.~J.}\ \bibnamefont {Holland}},\ }\bibfield  {title} {\bibinfo {title} {Atomtronics: Ultracold-atom analogs of electronic devices},\ }\href {https://doi.org/10.1103/PhysRevA.75.023615} {\bibfield  {journal} {\bibinfo  {journal} {Phys. Rev. A}\ }\textbf {\bibinfo {volume} {75}},\ \bibinfo {pages} {023615} (\bibinfo {year} {2007})}\BibitemShut {NoStop}%
\bibitem [{\citenamefont {Amico}\ \emph {et~al.}(2014)\citenamefont {Amico}, \citenamefont {Aghamalyan}, \citenamefont {Auksztol}, \citenamefont {Crepaz}, \citenamefont {Dumke},\ and\ \citenamefont {Kwek}}]{Amico2014}%
  \BibitemOpen
  \bibfield  {author} {\bibinfo {author} {\bibfnamefont {L.}~\bibnamefont {Amico}}, \bibinfo {author} {\bibfnamefont {D.}~\bibnamefont {Aghamalyan}}, \bibinfo {author} {\bibfnamefont {F.}~\bibnamefont {Auksztol}}, \bibinfo {author} {\bibfnamefont {H.}~\bibnamefont {Crepaz}}, \bibinfo {author} {\bibfnamefont {R.}~\bibnamefont {Dumke}},\ and\ \bibinfo {author} {\bibfnamefont {L.~C.}\ \bibnamefont {Kwek}},\ }\bibfield  {title} {\bibinfo {title} {Superfluid qubit systems with ring shaped optical lattices},\ }\href {https://doi.org/10.1038/srep04298} {\bibfield  {journal} {\bibinfo  {journal} {Scientific Reports}\ }\textbf {\bibinfo {volume} {4}},\ \bibinfo {pages} {4298} (\bibinfo {year} {2014})}\BibitemShut {NoStop}%
\bibitem [{\citenamefont {Micheli}\ \emph {et~al.}(2004)\citenamefont {Micheli}, \citenamefont {Daley}, \citenamefont {Jaksch},\ and\ \citenamefont {Zoller}}]{Micheli2004}%
  \BibitemOpen
  \bibfield  {author} {\bibinfo {author} {\bibfnamefont {A.}~\bibnamefont {Micheli}}, \bibinfo {author} {\bibfnamefont {A.~J.}\ \bibnamefont {Daley}}, \bibinfo {author} {\bibfnamefont {D.}~\bibnamefont {Jaksch}},\ and\ \bibinfo {author} {\bibfnamefont {P.}~\bibnamefont {Zoller}},\ }\bibfield  {title} {\bibinfo {title} {{Single Atom Transistor in a 1D Optical Lattice}},\ }\href {https://doi.org/10.1103/PhysRevLett.93.140408} {\bibfield  {journal} {\bibinfo  {journal} {Phys. Rev. Lett.}\ }\textbf {\bibinfo {volume} {93}},\ \bibinfo {pages} {140408} (\bibinfo {year} {2004})}\BibitemShut {NoStop}%
\bibitem [{\citenamefont {Stickney}\ \emph {et~al.}(2007)\citenamefont {Stickney}, \citenamefont {Anderson},\ and\ \citenamefont {Zozulya}}]{Stickney2007}%
  \BibitemOpen
  \bibfield  {author} {\bibinfo {author} {\bibfnamefont {J.~A.}\ \bibnamefont {Stickney}}, \bibinfo {author} {\bibfnamefont {D.~Z.}\ \bibnamefont {Anderson}},\ and\ \bibinfo {author} {\bibfnamefont {A.~A.}\ \bibnamefont {Zozulya}},\ }\bibfield  {title} {\bibinfo {title} {{Transistorlike behavior of a Bose-Einstein condensate in a triple-well potential}},\ }\href {https://doi.org/10.1103/PhysRevA.75.013608} {\bibfield  {journal} {\bibinfo  {journal} {Phys. Rev. A}\ }\textbf {\bibinfo {volume} {75}},\ \bibinfo {pages} {013608} (\bibinfo {year} {2007})}\BibitemShut {NoStop}%
\bibitem [{\citenamefont {Ruschhaupt}\ and\ \citenamefont {Muga}(2007)}]{Ruschhaupt2007}%
  \BibitemOpen
  \bibfield  {author} {\bibinfo {author} {\bibfnamefont {A.}~\bibnamefont {Ruschhaupt}}\ and\ \bibinfo {author} {\bibfnamefont {J.~G.}\ \bibnamefont {Muga}},\ }\bibfield  {title} {\bibinfo {title} {{Three-dimensional effects in atom diodes: Atom-optical devices for one-way motion}},\ }\href {https://doi.org/10.1103/PhysRevA.76.013619} {\bibfield  {journal} {\bibinfo  {journal} {Phys. Rev. A}\ }\textbf {\bibinfo {volume} {76}},\ \bibinfo {pages} {013619} (\bibinfo {year} {2007})}\BibitemShut {NoStop}%
\bibitem [{\citenamefont {Thorn}\ \emph {et~al.}(2008)\citenamefont {Thorn}, \citenamefont {Schoene}, \citenamefont {Li},\ and\ \citenamefont {Steck}}]{Thorn2008}%
  \BibitemOpen
  \bibfield  {author} {\bibinfo {author} {\bibfnamefont {J.~J.}\ \bibnamefont {Thorn}}, \bibinfo {author} {\bibfnamefont {E.~A.}\ \bibnamefont {Schoene}}, \bibinfo {author} {\bibfnamefont {T.}~\bibnamefont {Li}},\ and\ \bibinfo {author} {\bibfnamefont {D.~A.}\ \bibnamefont {Steck}},\ }\bibfield  {title} {\bibinfo {title} {Experimental realization of an optical one-way barrier for neutral atoms},\ }\href {https://doi.org/10.1103/PhysRevLett.100.240407} {\bibfield  {journal} {\bibinfo  {journal} {Phys. Rev. Lett.}\ }\textbf {\bibinfo {volume} {100}},\ \bibinfo {pages} {240407} (\bibinfo {year} {2008})}\BibitemShut {NoStop}%
\bibitem [{\citenamefont {Pepino}\ \emph {et~al.}(2009)\citenamefont {Pepino}, \citenamefont {Cooper}, \citenamefont {Anderson},\ and\ \citenamefont {Holland}}]{Pepino2009}%
  \BibitemOpen
  \bibfield  {author} {\bibinfo {author} {\bibfnamefont {R.~A.}\ \bibnamefont {Pepino}}, \bibinfo {author} {\bibfnamefont {J.}~\bibnamefont {Cooper}}, \bibinfo {author} {\bibfnamefont {D.~Z.}\ \bibnamefont {Anderson}},\ and\ \bibinfo {author} {\bibfnamefont {M.~J.}\ \bibnamefont {Holland}},\ }\bibfield  {title} {\bibinfo {title} {{Atomtronic Circuits of Diodes and Transistors}},\ }\href {https://doi.org/10.1103/PhysRevLett.103.140405} {\bibfield  {journal} {\bibinfo  {journal} {Phys. Rev. Lett.}\ }\textbf {\bibinfo {volume} {103}},\ \bibinfo {pages} {140405} (\bibinfo {year} {2009})}\BibitemShut {NoStop}%
\bibitem [{\citenamefont {Krinner}\ \emph {et~al.}(2017)\citenamefont {Krinner}, \citenamefont {Esslinger},\ and\ \citenamefont {Brantut}}]{Krinner2017}%
  \BibitemOpen
  \bibfield  {author} {\bibinfo {author} {\bibfnamefont {S.}~\bibnamefont {Krinner}}, \bibinfo {author} {\bibfnamefont {T.}~\bibnamefont {Esslinger}},\ and\ \bibinfo {author} {\bibfnamefont {J.-P.}\ \bibnamefont {Brantut}},\ }\bibfield  {title} {\bibinfo {title} {Two-terminal transport measurements with cold atoms},\ }\href {https://doi.org/10.1088/1361-648X/aa74a1} {\bibfield  {journal} {\bibinfo  {journal} {Journal of Physics: Condensed Matter}\ }\textbf {\bibinfo {volume} {29}},\ \bibinfo {pages} {343003} (\bibinfo {year} {2017})}\BibitemShut {NoStop}%
\bibitem [{\citenamefont {Zozulya}\ and\ \citenamefont {Anderson}(2013)}]{Zozulya2013}%
  \BibitemOpen
  \bibfield  {author} {\bibinfo {author} {\bibfnamefont {A.~A.}\ \bibnamefont {Zozulya}}\ and\ \bibinfo {author} {\bibfnamefont {D.~Z.}\ \bibnamefont {Anderson}},\ }\bibfield  {title} {\bibinfo {title} {Principles of an atomtronic battery},\ }\href {https://doi.org/10.1103/PhysRevA.88.043641} {\bibfield  {journal} {\bibinfo  {journal} {Phys. Rev. A}\ }\textbf {\bibinfo {volume} {88}},\ \bibinfo {pages} {043641} (\bibinfo {year} {2013})}\BibitemShut {NoStop}%
\bibitem [{\citenamefont {Caliga}\ \emph {et~al.}(2017)\citenamefont {Caliga}, \citenamefont {Straatsma},\ and\ \citenamefont {Anderson}}]{Caliga2017}%
  \BibitemOpen
  \bibfield  {author} {\bibinfo {author} {\bibfnamefont {S.~C.}\ \bibnamefont {Caliga}}, \bibinfo {author} {\bibfnamefont {C.~J.~E.}\ \bibnamefont {Straatsma}},\ and\ \bibinfo {author} {\bibfnamefont {D.~Z.}\ \bibnamefont {Anderson}},\ }\bibfield  {title} {\bibinfo {title} {Experimental demonstration of an atomtronic battery},\ }\href {https://doi.org/10.1088/1367-2630/aa56d8} {\bibfield  {journal} {\bibinfo  {journal} {New Journal of Physics}\ }\textbf {\bibinfo {volume} {19}},\ \bibinfo {pages} {013036} (\bibinfo {year} {2017})}\BibitemShut {NoStop}%
\bibitem [{\citenamefont {Caliga}\ \emph {et~al.}(2016)\citenamefont {Caliga}, \citenamefont {Straatsma}, \citenamefont {Zozulya},\ and\ \citenamefont {Anderson}}]{Caliga2016}%
  \BibitemOpen
  \bibfield  {author} {\bibinfo {author} {\bibfnamefont {S.~C.}\ \bibnamefont {Caliga}}, \bibinfo {author} {\bibfnamefont {C.~J.~E.}\ \bibnamefont {Straatsma}}, \bibinfo {author} {\bibfnamefont {A.~A.}\ \bibnamefont {Zozulya}},\ and\ \bibinfo {author} {\bibfnamefont {D.~Z.}\ \bibnamefont {Anderson}},\ }\bibfield  {title} {\bibinfo {title} {Principles of an atomtronic transistor},\ }\href {https://doi.org/10.1088/1367-2630/18/1/015012} {\bibfield  {journal} {\bibinfo  {journal} {New Journal of Physics}\ }\textbf {\bibinfo {volume} {18}},\ \bibinfo {pages} {015012} (\bibinfo {year} {2016})}\BibitemShut {NoStop}%
\bibitem [{\citenamefont {J{\"a}hrling}\ \emph {et~al.}(2024)\citenamefont {J{\"a}hrling}, \citenamefont {Singh},\ and\ \citenamefont {Mathey}}]{Jahrling2024}%
  \BibitemOpen
  \bibfield  {author} {\bibinfo {author} {\bibfnamefont {S.}~\bibnamefont {J{\"a}hrling}}, \bibinfo {author} {\bibfnamefont {V.~P.}\ \bibnamefont {Singh}},\ and\ \bibinfo {author} {\bibfnamefont {L.}~\bibnamefont {Mathey}},\ }\href@noop {} {\bibinfo {title} {{Designing Atomtronic Circuits via Superfluid Dynamics}}} (\bibinfo {year} {2024}),\ \Eprint {https://arxiv.org/abs/arXiv:2411.13642} {arXiv:2411.13642} \BibitemShut {NoStop}%
\bibitem [{\citenamefont {Albiez}\ \emph {et~al.}(2005)\citenamefont {Albiez}, \citenamefont {Gati}, \citenamefont {F\"olling}, \citenamefont {Hunsmann}, \citenamefont {Cristiani},\ and\ \citenamefont {Oberthaler}}]{Albiez2005}%
  \BibitemOpen
  \bibfield  {author} {\bibinfo {author} {\bibfnamefont {M.}~\bibnamefont {Albiez}}, \bibinfo {author} {\bibfnamefont {R.}~\bibnamefont {Gati}}, \bibinfo {author} {\bibfnamefont {J.}~\bibnamefont {F\"olling}}, \bibinfo {author} {\bibfnamefont {S.}~\bibnamefont {Hunsmann}}, \bibinfo {author} {\bibfnamefont {M.}~\bibnamefont {Cristiani}},\ and\ \bibinfo {author} {\bibfnamefont {M.~K.}\ \bibnamefont {Oberthaler}},\ }\bibfield  {title} {\bibinfo {title} {{Direct Observation of Tunneling and Nonlinear Self-Trapping in a Single Bosonic Josephson Junction}},\ }\href {https://doi.org/10.1103/PhysRevLett.95.010402} {\bibfield  {journal} {\bibinfo  {journal} {Phys. Rev. Lett.}\ }\textbf {\bibinfo {volume} {95}},\ \bibinfo {pages} {010402} (\bibinfo {year} {2005})}\BibitemShut {NoStop}%
\bibitem [{\citenamefont {Levy}\ \emph {et~al.}(2007)\citenamefont {Levy}, \citenamefont {Lahoud}, \citenamefont {Shomroni},\ and\ \citenamefont {Steinhauer}}]{Levy2007}%
  \BibitemOpen
  \bibfield  {author} {\bibinfo {author} {\bibfnamefont {S.}~\bibnamefont {Levy}}, \bibinfo {author} {\bibfnamefont {E.}~\bibnamefont {Lahoud}}, \bibinfo {author} {\bibfnamefont {I.}~\bibnamefont {Shomroni}},\ and\ \bibinfo {author} {\bibfnamefont {J.}~\bibnamefont {Steinhauer}},\ }\bibfield  {title} {\bibinfo {title} {{The a.c. and d.c. Josephson effects in a Bose--Einstein condensate}},\ }\href {https://doi.org/10.1038/nature06186} {\bibfield  {journal} {\bibinfo  {journal} {Nature}\ }\textbf {\bibinfo {volume} {449}},\ \bibinfo {pages} {579} (\bibinfo {year} {2007})}\BibitemShut {NoStop}%
\bibitem [{\citenamefont {Luick}\ \emph {et~al.}(2020)\citenamefont {Luick}, \citenamefont {Sobirey}, \citenamefont {Bohlen}, \citenamefont {Singh}, \citenamefont {Mathey}, \citenamefont {Lompe},\ and\ \citenamefont {Moritz}}]{Luick2020}%
  \BibitemOpen
  \bibfield  {author} {\bibinfo {author} {\bibfnamefont {N.}~\bibnamefont {Luick}}, \bibinfo {author} {\bibfnamefont {L.}~\bibnamefont {Sobirey}}, \bibinfo {author} {\bibfnamefont {M.}~\bibnamefont {Bohlen}}, \bibinfo {author} {\bibfnamefont {V.~P.}\ \bibnamefont {Singh}}, \bibinfo {author} {\bibfnamefont {L.}~\bibnamefont {Mathey}}, \bibinfo {author} {\bibfnamefont {T.}~\bibnamefont {Lompe}},\ and\ \bibinfo {author} {\bibfnamefont {H.}~\bibnamefont {Moritz}},\ }\bibfield  {title} {\bibinfo {title} {{An ideal Josephson junction in an ultracold two-dimensional Fermi gas}},\ }\href {https://doi.org/10.1126/science.aaz2342} {\bibfield  {journal} {\bibinfo  {journal} {Science}\ }\textbf {\bibinfo {volume} {369}},\ \bibinfo {pages} {89} (\bibinfo {year} {2020})}\BibitemShut {NoStop}%
\bibitem [{\citenamefont {Kwon}\ \emph {et~al.}(2020)\citenamefont {Kwon}, \citenamefont {Pace}, \citenamefont {Panza}, \citenamefont {Inguscio}, \citenamefont {Zwerger}, \citenamefont {Zaccanti}, \citenamefont {Scazza},\ and\ \citenamefont {Roati}}]{Kwon2020}%
  \BibitemOpen
  \bibfield  {author} {\bibinfo {author} {\bibfnamefont {W.~J.}\ \bibnamefont {Kwon}}, \bibinfo {author} {\bibfnamefont {G.~D.}\ \bibnamefont {Pace}}, \bibinfo {author} {\bibfnamefont {R.}~\bibnamefont {Panza}}, \bibinfo {author} {\bibfnamefont {M.}~\bibnamefont {Inguscio}}, \bibinfo {author} {\bibfnamefont {W.}~\bibnamefont {Zwerger}}, \bibinfo {author} {\bibfnamefont {M.}~\bibnamefont {Zaccanti}}, \bibinfo {author} {\bibfnamefont {F.}~\bibnamefont {Scazza}},\ and\ \bibinfo {author} {\bibfnamefont {G.}~\bibnamefont {Roati}},\ }\bibfield  {title} {\bibinfo {title} {{Strongly correlated superfluid order parameters from dc Josephson supercurrents}},\ }\href {https://doi.org/10.1126/science.aaz2463} {\bibfield  {journal} {\bibinfo  {journal} {Science}\ }\textbf {\bibinfo {volume} {369}},\ \bibinfo {pages} {84} (\bibinfo {year} {2020})}\BibitemShut {NoStop}%
\bibitem [{\citenamefont {Del~Pace}\ \emph {et~al.}(2021)\citenamefont {Del~Pace}, \citenamefont {Kwon}, \citenamefont {Zaccanti}, \citenamefont {Roati},\ and\ \citenamefont {Scazza}}]{Pace2021}%
  \BibitemOpen
  \bibfield  {author} {\bibinfo {author} {\bibfnamefont {G.}~\bibnamefont {Del~Pace}}, \bibinfo {author} {\bibfnamefont {W.~J.}\ \bibnamefont {Kwon}}, \bibinfo {author} {\bibfnamefont {M.}~\bibnamefont {Zaccanti}}, \bibinfo {author} {\bibfnamefont {G.}~\bibnamefont {Roati}},\ and\ \bibinfo {author} {\bibfnamefont {F.}~\bibnamefont {Scazza}},\ }\bibfield  {title} {\bibinfo {title} {{Tunneling Transport of Unitary Fermions across the Superfluid Transition}},\ }\href {https://doi.org/10.1103/PhysRevLett.126.055301} {\bibfield  {journal} {\bibinfo  {journal} {Phys. Rev. Lett.}\ }\textbf {\bibinfo {volume} {126}},\ \bibinfo {pages} {055301} (\bibinfo {year} {2021})}\BibitemShut {NoStop}%
\bibitem [{\citenamefont {Singh}\ \emph {et~al.}(2024)\citenamefont {Singh}, \citenamefont {Polo}, \citenamefont {Mathey},\ and\ \citenamefont {Amico}}]{SinghShapiro}%
  \BibitemOpen
  \bibfield  {author} {\bibinfo {author} {\bibfnamefont {V.~P.}\ \bibnamefont {Singh}}, \bibinfo {author} {\bibfnamefont {J.}~\bibnamefont {Polo}}, \bibinfo {author} {\bibfnamefont {L.}~\bibnamefont {Mathey}},\ and\ \bibinfo {author} {\bibfnamefont {L.}~\bibnamefont {Amico}},\ }\bibfield  {title} {\bibinfo {title} {{Shapiro Steps in Driven Atomic Josephson Junctions}},\ }\href {https://doi.org/10.1103/PhysRevLett.133.093401} {\bibfield  {journal} {\bibinfo  {journal} {Phys. Rev. Lett.}\ }\textbf {\bibinfo {volume} {133}},\ \bibinfo {pages} {093401} (\bibinfo {year} {2024})}\BibitemShut {NoStop}%
\bibitem [{\citenamefont {Del~Pace}\ \emph {et~al.}(2024)\citenamefont {Del~Pace}, \citenamefont {Hern{\'a}ndez-Rajkov}, \citenamefont {Singh}, \citenamefont {Grani}, \citenamefont {Fern{\'a}ndez}, \citenamefont {Nesti}, \citenamefont {Seman}, \citenamefont {Inguscio}, \citenamefont {Amico},\ and\ \citenamefont {Roati}}]{Del_Pace2024}%
  \BibitemOpen
  \bibfield  {author} {\bibinfo {author} {\bibfnamefont {G.}~\bibnamefont {Del~Pace}}, \bibinfo {author} {\bibfnamefont {D.}~\bibnamefont {Hern{\'a}ndez-Rajkov}}, \bibinfo {author} {\bibfnamefont {V.~P.}\ \bibnamefont {Singh}}, \bibinfo {author} {\bibfnamefont {N.}~\bibnamefont {Grani}}, \bibinfo {author} {\bibfnamefont {M.~F.}\ \bibnamefont {Fern{\'a}ndez}}, \bibinfo {author} {\bibfnamefont {G.}~\bibnamefont {Nesti}}, \bibinfo {author} {\bibfnamefont {J.~A.}\ \bibnamefont {Seman}}, \bibinfo {author} {\bibfnamefont {M.}~\bibnamefont {Inguscio}}, \bibinfo {author} {\bibfnamefont {L.}~\bibnamefont {Amico}},\ and\ \bibinfo {author} {\bibfnamefont {G.}~\bibnamefont {Roati}},\ }\bibfield  {title} {\bibinfo {title} {{Shapiro steps in strongly-interacting Fermi gases}},\ }\href@noop {} {\  (\bibinfo {year} {2024})},\ \Eprint {https://arxiv.org/abs/2409.03448} {arXiv:2409.03448 [cond-mat.quant-gas]} \BibitemShut {NoStop}%
\bibitem [{\citenamefont {Bernhart}\ \emph {et~al.}(2024)\citenamefont {Bernhart}, \citenamefont {R{\"o}hrle}, \citenamefont {Singh}, \citenamefont {Mathey}, \citenamefont {Amico},\ and\ \citenamefont {Ott}}]{Bernhart2024}%
  \BibitemOpen
  \bibfield  {author} {\bibinfo {author} {\bibfnamefont {E.}~\bibnamefont {Bernhart}}, \bibinfo {author} {\bibfnamefont {M.}~\bibnamefont {R{\"o}hrle}}, \bibinfo {author} {\bibfnamefont {V.~P.}\ \bibnamefont {Singh}}, \bibinfo {author} {\bibfnamefont {L.}~\bibnamefont {Mathey}}, \bibinfo {author} {\bibfnamefont {L.}~\bibnamefont {Amico}},\ and\ \bibinfo {author} {\bibfnamefont {H.}~\bibnamefont {Ott}},\ }\bibfield  {title} {\bibinfo {title} {{Observation of Shapiro steps in an ultracold atomic Josephson junction}},\ }\href@noop {} {\  (\bibinfo {year} {2024})},\ \Eprint {https://arxiv.org/abs/2409.03340} {arXiv:2409.03340 [cond-mat.quant-gas]} \BibitemShut {NoStop}%
\bibitem [{\citenamefont {Singh}\ \emph {et~al.}(2025)\citenamefont {Singh}, \citenamefont {Amico},\ and\ \citenamefont {Mathey}}]{Singh2025}%
  \BibitemOpen
  \bibfield  {author} {\bibinfo {author} {\bibfnamefont {V.~P.}\ \bibnamefont {Singh}}, \bibinfo {author} {\bibfnamefont {L.}~\bibnamefont {Amico}},\ and\ \bibinfo {author} {\bibfnamefont {L.}~\bibnamefont {Mathey}},\ }\href@noop {} {\bibinfo {title} {{Atomic Josephson Parametric Amplifier}}} (\bibinfo {year} {2025}),\ \Eprint {https://arxiv.org/abs/arXiv:2503.20890} {arXiv:2503.20890} \BibitemShut {NoStop}%
\bibitem [{\citenamefont {Gross}\ \emph {et~al.}(2016)\citenamefont {Gross}, \citenamefont {Marx},\ and\ \citenamefont {Deppe}}]{gross2016}%
  \BibitemOpen
  \bibfield  {author} {\bibinfo {author} {\bibfnamefont {R.}~\bibnamefont {Gross}}, \bibinfo {author} {\bibfnamefont {A.}~\bibnamefont {Marx}},\ and\ \bibinfo {author} {\bibfnamefont {F.}~\bibnamefont {Deppe}},\ }\href {https://books.google.de/books?id=4SIzrgEACAAJ} {\emph {\bibinfo {title} {Applied Superconductivity: Josephson Effect and Superconducting Electronics}}},\ De Gruyter Textbook\ (\bibinfo  {publisher} {De Gruyter},\ \bibinfo {year} {2016})\BibitemShut {NoStop}%
\bibitem [{\citenamefont {Vrba}\ \emph {et~al.}(2006)\citenamefont {Vrba}, \citenamefont {Nenonen},\ and\ \citenamefont {Trahms}}]{vrba2006}%
  \BibitemOpen
  \bibfield  {author} {\bibinfo {author} {\bibfnamefont {J.}~\bibnamefont {Vrba}}, \bibinfo {author} {\bibfnamefont {J.}~\bibnamefont {Nenonen}},\ and\ \bibinfo {author} {\bibfnamefont {L.}~\bibnamefont {Trahms}},\ }\bibinfo {title} {Biomagnetism},\ in\ \href {https://doi.org/https://doi.org/10.1002/9783527609956.ch11} {\emph {\bibinfo {booktitle} {The SQUID Handbook}}}\ (\bibinfo  {publisher} {John Wiley \& Sons, Ltd},\ \bibinfo {year} {2006})\ Chap.~\bibinfo {chapter} {11}, pp.\ \bibinfo {pages} {269--389}\BibitemShut {NoStop}%
\bibitem [{\citenamefont {Ramanathan}\ \emph {et~al.}(2011)\citenamefont {Ramanathan}, \citenamefont {Wright}, \citenamefont {Muniz}, \citenamefont {Zelan}, \citenamefont {Hill}, \citenamefont {Lobb}, \citenamefont {Helmerson}, \citenamefont {Phillips},\ and\ \citenamefont {Campbell}}]{Ramanathan2011}%
  \BibitemOpen
  \bibfield  {author} {\bibinfo {author} {\bibfnamefont {A.}~\bibnamefont {Ramanathan}}, \bibinfo {author} {\bibfnamefont {K.~C.}\ \bibnamefont {Wright}}, \bibinfo {author} {\bibfnamefont {S.~R.}\ \bibnamefont {Muniz}}, \bibinfo {author} {\bibfnamefont {M.}~\bibnamefont {Zelan}}, \bibinfo {author} {\bibfnamefont {W.~T.}\ \bibnamefont {Hill}}, \bibinfo {author} {\bibfnamefont {C.~J.}\ \bibnamefont {Lobb}}, \bibinfo {author} {\bibfnamefont {K.}~\bibnamefont {Helmerson}}, \bibinfo {author} {\bibfnamefont {W.~D.}\ \bibnamefont {Phillips}},\ and\ \bibinfo {author} {\bibfnamefont {G.~K.}\ \bibnamefont {Campbell}},\ }\bibfield  {title} {\bibinfo {title} {{Superflow in a Toroidal Bose-Einstein Condensate: An Atom Circuit with a Tunable Weak Link}},\ }\href {https://doi.org/10.1103/PhysRevLett.106.130401} {\bibfield  {journal} {\bibinfo  {journal} {Phys. Rev. Lett.}\ }\textbf {\bibinfo {volume} {106}},\ \bibinfo {pages} {130401} (\bibinfo {year} {2011})}\BibitemShut {NoStop}%
\bibitem [{\citenamefont {Del~Pace}\ \emph {et~al.}(2022)\citenamefont {Del~Pace}, \citenamefont {Xhani}, \citenamefont {Muzi~Falconi}, \citenamefont {Fedrizzi}, \citenamefont {Grani}, \citenamefont {Hernandez~Rajkov}, \citenamefont {Inguscio}, \citenamefont {Scazza}, \citenamefont {Kwon},\ and\ \citenamefont {Roati}}]{Pace2022}%
  \BibitemOpen
  \bibfield  {author} {\bibinfo {author} {\bibfnamefont {G.}~\bibnamefont {Del~Pace}}, \bibinfo {author} {\bibfnamefont {K.}~\bibnamefont {Xhani}}, \bibinfo {author} {\bibfnamefont {A.}~\bibnamefont {Muzi~Falconi}}, \bibinfo {author} {\bibfnamefont {M.}~\bibnamefont {Fedrizzi}}, \bibinfo {author} {\bibfnamefont {N.}~\bibnamefont {Grani}}, \bibinfo {author} {\bibfnamefont {D.}~\bibnamefont {Hernandez~Rajkov}}, \bibinfo {author} {\bibfnamefont {M.}~\bibnamefont {Inguscio}}, \bibinfo {author} {\bibfnamefont {F.}~\bibnamefont {Scazza}}, \bibinfo {author} {\bibfnamefont {W.~J.}\ \bibnamefont {Kwon}},\ and\ \bibinfo {author} {\bibfnamefont {G.}~\bibnamefont {Roati}},\ }\bibfield  {title} {\bibinfo {title} {{Imprinting Persistent Currents in Tunable Fermionic Rings}},\ }\href {https://doi.org/10.1103/PhysRevX.12.041037} {\bibfield  {journal} {\bibinfo  {journal} {Phys. Rev. X}\ }\textbf {\bibinfo {volume} {12}},\ \bibinfo {pages} {041037} (\bibinfo {year} {2022})}\BibitemShut {NoStop}%
\bibitem [{\citenamefont {Cai}\ \emph {et~al.}(2022)\citenamefont {Cai}, \citenamefont {Allman}, \citenamefont {Sabharwal},\ and\ \citenamefont {Wright}}]{Cai2022}%
  \BibitemOpen
  \bibfield  {author} {\bibinfo {author} {\bibfnamefont {Y.}~\bibnamefont {Cai}}, \bibinfo {author} {\bibfnamefont {D.~G.}\ \bibnamefont {Allman}}, \bibinfo {author} {\bibfnamefont {P.}~\bibnamefont {Sabharwal}},\ and\ \bibinfo {author} {\bibfnamefont {K.~C.}\ \bibnamefont {Wright}},\ }\bibfield  {title} {\bibinfo {title} {{Persistent Currents in Rings of Ultracold Fermionic Atoms}},\ }\href {https://doi.org/10.1103/PhysRevLett.128.150401} {\bibfield  {journal} {\bibinfo  {journal} {Phys. Rev. Lett.}\ }\textbf {\bibinfo {volume} {128}},\ \bibinfo {pages} {150401} (\bibinfo {year} {2022})}\BibitemShut {NoStop}%
\bibitem [{\citenamefont {Mathey}\ \emph {et~al.}(2014)\citenamefont {Mathey}, \citenamefont {Clark},\ and\ \citenamefont {Mathey}}]{Mathey2014}%
  \BibitemOpen
  \bibfield  {author} {\bibinfo {author} {\bibfnamefont {A.~C.}\ \bibnamefont {Mathey}}, \bibinfo {author} {\bibfnamefont {C.~W.}\ \bibnamefont {Clark}},\ and\ \bibinfo {author} {\bibfnamefont {L.}~\bibnamefont {Mathey}},\ }\bibfield  {title} {\bibinfo {title} {Decay of a superfluid current of ultracold atoms in a toroidal trap},\ }\href {https://doi.org/10.1103/PhysRevA.90.023604} {\bibfield  {journal} {\bibinfo  {journal} {Phys. Rev. A}\ }\textbf {\bibinfo {volume} {90}},\ \bibinfo {pages} {023604} (\bibinfo {year} {2014})}\BibitemShut {NoStop}%
\bibitem [{\citenamefont {Polo}\ \emph {et~al.}(2025)\citenamefont {Polo}, \citenamefont {Chetcuti}, \citenamefont {Haug}, \citenamefont {Minguzzi}, \citenamefont {Wright},\ and\ \citenamefont {Amico}}]{POLO2025}%
  \BibitemOpen
  \bibfield  {author} {\bibinfo {author} {\bibfnamefont {J.}~\bibnamefont {Polo}}, \bibinfo {author} {\bibfnamefont {W.}~\bibnamefont {Chetcuti}}, \bibinfo {author} {\bibfnamefont {T.}~\bibnamefont {Haug}}, \bibinfo {author} {\bibfnamefont {A.}~\bibnamefont {Minguzzi}}, \bibinfo {author} {\bibfnamefont {K.}~\bibnamefont {Wright}},\ and\ \bibinfo {author} {\bibfnamefont {L.}~\bibnamefont {Amico}},\ }\bibfield  {title} {\bibinfo {title} {Persistent currents in ultracold gases},\ }\href {https://doi.org/https://doi.org/10.1016/j.physrep.2025.06.003} {\bibfield  {journal} {\bibinfo  {journal} {Physics Reports}\ }\textbf {\bibinfo {volume} {1137}},\ \bibinfo {pages} {1} (\bibinfo {year} {2025})}\BibitemShut {NoStop}%
\bibitem [{\citenamefont {Wright}\ \emph {et~al.}(2013)\citenamefont {Wright}, \citenamefont {Blakestad}, \citenamefont {Lobb}, \citenamefont {Phillips},\ and\ \citenamefont {Campbell}}]{Wright2013}%
  \BibitemOpen
  \bibfield  {author} {\bibinfo {author} {\bibfnamefont {K.~C.}\ \bibnamefont {Wright}}, \bibinfo {author} {\bibfnamefont {R.~B.}\ \bibnamefont {Blakestad}}, \bibinfo {author} {\bibfnamefont {C.~J.}\ \bibnamefont {Lobb}}, \bibinfo {author} {\bibfnamefont {W.~D.}\ \bibnamefont {Phillips}},\ and\ \bibinfo {author} {\bibfnamefont {G.~K.}\ \bibnamefont {Campbell}},\ }\bibfield  {title} {\bibinfo {title} {{Driving Phase Slips in a Superfluid Atom Circuit with a Rotating Weak Link}},\ }\href {https://doi.org/10.1103/PhysRevLett.110.025302} {\bibfield  {journal} {\bibinfo  {journal} {Phys. Rev. Lett.}\ }\textbf {\bibinfo {volume} {110}},\ \bibinfo {pages} {025302} (\bibinfo {year} {2013})}\BibitemShut {NoStop}%
\bibitem [{\citenamefont {Ryu}\ \emph {et~al.}(2013)\citenamefont {Ryu}, \citenamefont {Blackburn}, \citenamefont {Blinova},\ and\ \citenamefont {Boshier}}]{ryu2013}%
  \BibitemOpen
  \bibfield  {author} {\bibinfo {author} {\bibfnamefont {C.}~\bibnamefont {Ryu}}, \bibinfo {author} {\bibfnamefont {P.~W.}\ \bibnamefont {Blackburn}}, \bibinfo {author} {\bibfnamefont {A.~A.}\ \bibnamefont {Blinova}},\ and\ \bibinfo {author} {\bibfnamefont {M.~G.}\ \bibnamefont {Boshier}},\ }\bibfield  {title} {\bibinfo {title} {Experimental realization of josephson junctions for an atom squid},\ }\href {https://doi.org/10.1103/PhysRevLett.111.205301} {\bibfield  {journal} {\bibinfo  {journal} {Phys. Rev. Lett.}\ }\textbf {\bibinfo {volume} {111}},\ \bibinfo {pages} {205301} (\bibinfo {year} {2013})}\BibitemShut {NoStop}%
\bibitem [{\citenamefont {Eckel}\ \emph {et~al.}(2014)\citenamefont {Eckel}, \citenamefont {Lee}, \citenamefont {Jendrzejewski}, \citenamefont {Murray}, \citenamefont {Clark}, \citenamefont {Lobb}, \citenamefont {Phillips}, \citenamefont {Edwards},\ and\ \citenamefont {Campbell}}]{Eckel2014}%
  \BibitemOpen
  \bibfield  {author} {\bibinfo {author} {\bibfnamefont {S.}~\bibnamefont {Eckel}}, \bibinfo {author} {\bibfnamefont {J.~G.}\ \bibnamefont {Lee}}, \bibinfo {author} {\bibfnamefont {F.}~\bibnamefont {Jendrzejewski}}, \bibinfo {author} {\bibfnamefont {N.}~\bibnamefont {Murray}}, \bibinfo {author} {\bibfnamefont {C.~W.}\ \bibnamefont {Clark}}, \bibinfo {author} {\bibfnamefont {C.~J.}\ \bibnamefont {Lobb}}, \bibinfo {author} {\bibfnamefont {W.~D.}\ \bibnamefont {Phillips}}, \bibinfo {author} {\bibfnamefont {M.}~\bibnamefont {Edwards}},\ and\ \bibinfo {author} {\bibfnamefont {G.~K.}\ \bibnamefont {Campbell}},\ }\bibfield  {title} {\bibinfo {title} {Hysteresis in a quantized superfluid `atomtronic' circuit},\ }\href {https://doi.org/10.1038/nature12958} {\bibfield  {journal} {\bibinfo  {journal} {Nature}\ }\textbf {\bibinfo {volume} {506}},\ \bibinfo {pages} {200} (\bibinfo {year} {2014})}\BibitemShut {NoStop}%
\bibitem [{\citenamefont {Jendrzejewski}\ \emph {et~al.}(2014)\citenamefont {Jendrzejewski}, \citenamefont {Eckel}, \citenamefont {Murray}, \citenamefont {Lanier}, \citenamefont {Edwards}, \citenamefont {Lobb},\ and\ \citenamefont {Campbell}}]{Campbell2014}%
  \BibitemOpen
  \bibfield  {author} {\bibinfo {author} {\bibfnamefont {F.}~\bibnamefont {Jendrzejewski}}, \bibinfo {author} {\bibfnamefont {S.}~\bibnamefont {Eckel}}, \bibinfo {author} {\bibfnamefont {N.}~\bibnamefont {Murray}}, \bibinfo {author} {\bibfnamefont {C.}~\bibnamefont {Lanier}}, \bibinfo {author} {\bibfnamefont {M.}~\bibnamefont {Edwards}}, \bibinfo {author} {\bibfnamefont {C.~J.}\ \bibnamefont {Lobb}},\ and\ \bibinfo {author} {\bibfnamefont {G.~K.}\ \bibnamefont {Campbell}},\ }\bibfield  {title} {\bibinfo {title} {{Resistive Flow in a Weakly Interacting Bose-Einstein Condensate}},\ }\href {https://doi.org/10.1103/PhysRevLett.113.045305} {\bibfield  {journal} {\bibinfo  {journal} {Phys. Rev. Lett.}\ }\textbf {\bibinfo {volume} {113}},\ \bibinfo {pages} {045305} (\bibinfo {year} {2014})}\BibitemShut {NoStop}%
\bibitem [{\citenamefont {Ryu}\ \emph {et~al.}(2020)\citenamefont {Ryu}, \citenamefont {Samson},\ and\ \citenamefont {Boshier}}]{ryu2020}%
  \BibitemOpen
  \bibfield  {author} {\bibinfo {author} {\bibfnamefont {C.}~\bibnamefont {Ryu}}, \bibinfo {author} {\bibfnamefont {E.~C.}\ \bibnamefont {Samson}},\ and\ \bibinfo {author} {\bibfnamefont {M.~G.}\ \bibnamefont {Boshier}},\ }\bibfield  {title} {\bibinfo {title} {Quantum interference of currents in an atomtronic squid},\ }\href {https://doi.org/10.1038/s41467-020-17185-6} {\bibfield  {journal} {\bibinfo  {journal} {Nature Communications}\ }\textbf {\bibinfo {volume} {11}},\ \bibinfo {pages} {3338} (\bibinfo {year} {2020})}\BibitemShut {NoStop}%
\bibitem [{\citenamefont {Mathey}\ and\ \citenamefont {Mathey}(2016)}]{mathey2016}%
  \BibitemOpen
  \bibfield  {author} {\bibinfo {author} {\bibfnamefont {A.~C.}\ \bibnamefont {Mathey}}\ and\ \bibinfo {author} {\bibfnamefont {L.}~\bibnamefont {Mathey}},\ }\bibfield  {title} {\bibinfo {title} {Realizing and optimizing an atomtronic {SQUID}},\ }\href {https://doi.org/10.1088/1367-2630/18/5/055016} {\bibfield  {journal} {\bibinfo  {journal} {New Journal of Physics}\ }\textbf {\bibinfo {volume} {18}},\ \bibinfo {pages} {055016} (\bibinfo {year} {2016})}\BibitemShut {NoStop}%
\bibitem [{\citenamefont {Kiehn}\ \emph {et~al.}(2022)\citenamefont {Kiehn}, \citenamefont {Singh},\ and\ \citenamefont {Mathey}}]{kiehn2022}%
  \BibitemOpen
  \bibfield  {author} {\bibinfo {author} {\bibfnamefont {H.}~\bibnamefont {Kiehn}}, \bibinfo {author} {\bibfnamefont {V.~P.}\ \bibnamefont {Singh}},\ and\ \bibinfo {author} {\bibfnamefont {L.}~\bibnamefont {Mathey}},\ }\bibfield  {title} {\bibinfo {title} {{Implementation of an atomtronic SQUID in a strongly confined toroidal condensate}},\ }\href {https://doi.org/10.1103/PhysRevResearch.4.033024} {\bibfield  {journal} {\bibinfo  {journal} {Phys. Rev. Res.}\ }\textbf {\bibinfo {volume} {4}},\ \bibinfo {pages} {033024} (\bibinfo {year} {2022})}\BibitemShut {NoStop}%
\bibitem [{\citenamefont {Blakie}\ \emph {et~al.}(2008)\citenamefont {Blakie}, \citenamefont {Bradley}, \citenamefont {Davis}, \citenamefont {Ballagh},\ and\ \citenamefont {Gardiner}}]{Blakie2008}%
  \BibitemOpen
  \bibfield  {author} {\bibinfo {author} {\bibfnamefont {P.~B.}\ \bibnamefont {Blakie}}, \bibinfo {author} {\bibfnamefont {A.~S.}\ \bibnamefont {Bradley}}, \bibinfo {author} {\bibfnamefont {M.~J.}\ \bibnamefont {Davis}}, \bibinfo {author} {\bibfnamefont {R.~J.}\ \bibnamefont {Ballagh}},\ and\ \bibinfo {author} {\bibfnamefont {C.~W.}\ \bibnamefont {Gardiner}},\ }\bibfield  {title} {\bibinfo {title} {{Dynamics and statistical mechanics of ultra-cold Bose gases using c-field techniques}},\ }\href {https://doi.org/10.1080/00018730802564254} {\bibfield  {journal} {\bibinfo  {journal} {Advances in Physics}\ }\textbf {\bibinfo {volume} {57}},\ \bibinfo {pages} {363} (\bibinfo {year} {2008})}\BibitemShut {NoStop}%
\bibitem [{\citenamefont {Polkovnikov}(2010)}]{Polkovnikov2010}%
  \BibitemOpen
  \bibfield  {author} {\bibinfo {author} {\bibfnamefont {A.}~\bibnamefont {Polkovnikov}},\ }\bibfield  {title} {\bibinfo {title} {Phase space representation of quantum dynamics},\ }\href {https://doi.org/https://doi.org/10.1016/j.aop.2010.02.006} {\bibfield  {journal} {\bibinfo  {journal} {Annals of Physics}\ }\textbf {\bibinfo {volume} {325}},\ \bibinfo {pages} {1790} (\bibinfo {year} {2010})}\BibitemShut {NoStop}%
\bibitem [{\citenamefont {Singh}\ \emph {et~al.}(2017)\citenamefont {Singh}, \citenamefont {Weitenberg}, \citenamefont {Dalibard},\ and\ \citenamefont {Mathey}}]{Singh2017}%
  \BibitemOpen
  \bibfield  {author} {\bibinfo {author} {\bibfnamefont {V.~P.}\ \bibnamefont {Singh}}, \bibinfo {author} {\bibfnamefont {C.}~\bibnamefont {Weitenberg}}, \bibinfo {author} {\bibfnamefont {J.}~\bibnamefont {Dalibard}},\ and\ \bibinfo {author} {\bibfnamefont {L.}~\bibnamefont {Mathey}},\ }\bibfield  {title} {\bibinfo {title} {{Superfluidity and relaxation dynamics of a laser-stirred two-dimensional Bose gas}},\ }\href {https://doi.org/10.1103/PhysRevA.95.043631} {\bibfield  {journal} {\bibinfo  {journal} {Phys. Rev. A}\ }\textbf {\bibinfo {volume} {95}},\ \bibinfo {pages} {043631} (\bibinfo {year} {2017})}\BibitemShut {NoStop}%
\bibitem [{\citenamefont {Singh}\ and\ \citenamefont {Mathey}(2021)}]{Singh2021}%
  \BibitemOpen
  \bibfield  {author} {\bibinfo {author} {\bibfnamefont {V.~P.}\ \bibnamefont {Singh}}\ and\ \bibinfo {author} {\bibfnamefont {L.}~\bibnamefont {Mathey}},\ }\bibfield  {title} {\bibinfo {title} {{Collective modes and superfluidity of a two-dimensional ultracold Bose gas}},\ }\href {https://doi.org/10.1103/PhysRevResearch.3.023112} {\bibfield  {journal} {\bibinfo  {journal} {Phys. Rev. Res.}\ }\textbf {\bibinfo {volume} {3}},\ \bibinfo {pages} {023112} (\bibinfo {year} {2021})}\BibitemShut {NoStop}%
\bibitem [{\citenamefont {Cash}\ and\ \citenamefont {Karp}(1990)}]{cash1990}%
  \BibitemOpen
  \bibfield  {author} {\bibinfo {author} {\bibfnamefont {J.~R.}\ \bibnamefont {Cash}}\ and\ \bibinfo {author} {\bibfnamefont {A.~H.}\ \bibnamefont {Karp}},\ }\bibfield  {title} {\bibinfo {title} {A variable order runge-kutta method for initial value problems with rapidly varying right-hand sides},\ }\href@noop {} {\bibfield  {journal} {\bibinfo  {journal} {ACM Transactions on Mathematical Software (TOMS)}\ }\textbf {\bibinfo {volume} {16}},\ \bibinfo {pages} {201} (\bibinfo {year} {1990})}\BibitemShut {NoStop}%
\bibitem [{\citenamefont {Pethick}\ and\ \citenamefont {Smith}(2008)}]{pethick2008bose}%
  \BibitemOpen
  \bibfield  {author} {\bibinfo {author} {\bibfnamefont {C.~J.}\ \bibnamefont {Pethick}}\ and\ \bibinfo {author} {\bibfnamefont {H.}~\bibnamefont {Smith}},\ }\href@noop {} {\emph {\bibinfo {title} {Bose--Einstein condensation in dilute gases}}}\ (\bibinfo  {publisher} {Cambridge university press},\ \bibinfo {year} {2008})\BibitemShut {NoStop}%
\bibitem [{\citenamefont {Singh}\ \emph {et~al.}(2020)\citenamefont {Singh}, \citenamefont {Luick}, \citenamefont {Sobirey},\ and\ \citenamefont {Mathey}}]{Singh2020jj}%
  \BibitemOpen
  \bibfield  {author} {\bibinfo {author} {\bibfnamefont {V.~P.}\ \bibnamefont {Singh}}, \bibinfo {author} {\bibfnamefont {N.}~\bibnamefont {Luick}}, \bibinfo {author} {\bibfnamefont {L.}~\bibnamefont {Sobirey}},\ and\ \bibinfo {author} {\bibfnamefont {L.}~\bibnamefont {Mathey}},\ }\bibfield  {title} {\bibinfo {title} {{Josephson junction dynamics in a two-dimensional ultracold Bose gas}},\ }\href {https://doi.org/10.1103/PhysRevResearch.2.033298} {\bibfield  {journal} {\bibinfo  {journal} {Phys. Rev. Res.}\ }\textbf {\bibinfo {volume} {2}},\ \bibinfo {pages} {033298} (\bibinfo {year} {2020})}\BibitemShut {NoStop}%
\bibitem [{\citenamefont {Piazza}\ \emph {et~al.}(2009)\citenamefont {Piazza}, \citenamefont {Collins},\ and\ \citenamefont {Smerzi}}]{Piazza2009}%
  \BibitemOpen
  \bibfield  {author} {\bibinfo {author} {\bibfnamefont {F.}~\bibnamefont {Piazza}}, \bibinfo {author} {\bibfnamefont {L.~A.}\ \bibnamefont {Collins}},\ and\ \bibinfo {author} {\bibfnamefont {A.}~\bibnamefont {Smerzi}},\ }\bibfield  {title} {\bibinfo {title} {{Vortex-induced phase-slip dissipation in a toroidal Bose-Einstein condensate flowing through a barrier}},\ }\href {https://doi.org/10.1103/PhysRevA.80.021601} {\bibfield  {journal} {\bibinfo  {journal} {Phys. Rev. A}\ }\textbf {\bibinfo {volume} {80}},\ \bibinfo {pages} {021601} (\bibinfo {year} {2009})}\BibitemShut {NoStop}%
\bibitem [{\citenamefont {Clarke}\ and\ \citenamefont {Braginski}(2006)}]{clarke2006squid}%
  \BibitemOpen
  \bibfield  {author} {\bibinfo {author} {\bibfnamefont {J.}~\bibnamefont {Clarke}}\ and\ \bibinfo {author} {\bibfnamefont {A.~I.}\ \bibnamefont {Braginski}},\ }\href@noop {} {\emph {\bibinfo {title} {The SQUID handbook: Applications of SQUIDs and SQUID systems}}}\ (\bibinfo  {publisher} {John Wiley \& Sons},\ \bibinfo {year} {2006})\BibitemShut {NoStop}%
\end{thebibliography}%

\end{document}